 \documentclass[10pt,preprint]{aastex}  
 \usepackage{natbib}
\def\ang{\AA}
\def\arcsec{\hbox{$^{\prime\prime}$}}

\def\gapprox{\lower.4ex\hbox{$\;\buildrel >\over{\scriptstyle\sim}\;$}}
\def\lapprox{\lower.4ex\hbox{$\;\buildrel <\over{\scriptstyle\sim}\;$}}

\shortauthors{ASCHWANDEN 2012}
\shorttitle{Spatio-Temporal Evolution of Solar Flares}

\begin{document}

\title{         The Spatio-Temporal Evolution of Solar Flares 
		Observed with AIA/SDO: Fractal Diffusion, Sub-Diffusion,
		or Logistic Growth ?}

\author{        Markus J. Aschwanden$^2$		 }

\affil{
		Lockheed Martin Advanced Technology Center,
                Solar \& Astrophysics Laboratory,
                Org. ADBS, Bldg.252,
                3251 Hanover St.,
                Palo Alto, CA 94304, USA;
                e-mail: aschwanden@lmsal.com}

\begin{abstract}
We explore the spatio-temporal evolution of solar flares by fitting
a radial expansion model $r(t)$ that consists of an exponentially
growing acceleration phase, followed by a deceleration phase that
is parameterized by the generalized diffusion function 
$r(t) \propto \kappa (t-t_1)^{\beta/2}$, which
includes the logistic growth limit ($\beta=0$), sub-diffusion
($\beta = 0-1$), classical diffusion ($\beta=1$), super-diffusion
($\beta = 1-2$), and the linear expansion limit ($\beta=2$). 
We analyze all M and X-class flares observed with GOES and AIA/SDO
during the first two years of the SDO mission, amounting to 155 events.
We find that most flares operate in the sub-diffusive regime
($\beta=0.53\pm0.27$), which we interpret in terms of anisotropic
chain reactions of intermittent magnetic reconnection episodes 
in a low plasma-$\beta$ corona. We find a mean propagation speed of
$v=15\pm12$ km s$^{-1}$, with maximum speeds of $v_{max}=80 \pm 85$
km s$^{-1}$ per flare, which is substantially slower than the 
sonic speeds expected for thermal diffusion of flare plasmas. 
The diffusive characteristics established here (for the first time
for solar flares) is consistent with the fractal-diffusive 
self-organized criticality (FD-SOC) model, which predicted
diffusive transport merely based on cellular automaton simulations. 
\end{abstract}


\section{INTRODUCTION}

Many physical processes can be characterized by their spatio-temporal 
evolution, which we simply define here as the temporal evolution or time
dependence of a spatial or geometric parameter, say $x(t)$, where the 
geometric parameter $x$ could be a spatial length scale, an area, a volume, 
a fractal dimension, or any combination of these. At the largest scales,
for instance, the Big Bang theory describes the evolution of the universe 
by the gradual expansion of its radius $r(t)$, which can be decelerating 
in a closed universe (Friedmann-Lema\^itre-Robertson-Walker model).
Recent supernova observations reveal an accelerating universe at 
$z \approx 0.5$ (Riess et al.~1998; Perlmutter et al.~1999), and a 
deceleration that preceded the current epoch of cosmic acceleration 
(Riess et al.~2004). 

Other dynamic processes that predict a specific spatio-temporal 
evolution based on some physical model include Brownian motion, 
fractal Brownian motion, classical diffusion, sub-diffusion, 
super-diffusion, L\'evy flights, logistic growth, percolation, 
self-organized criticality avalanches, cellular automatons, non-extensive 
Tsallis entropy, complex networks, etc. The knowledge of the 
spatio-temporal evolution of a physical process is often directly related 
to scaling laws between various physical parameters, and thus spatio-temporal
measurements play a decisive role in the derivation of scaling laws. 
For instance, particle conservation ($n V = const$) in an expanding gas, 
plasma, or universe predicts a reciprocal scaling between the particle 
density $n(t)$ and the volume $V(t)$, i.e., $n(t) \propto V(t)^{-1}$, 
and consequently a scaling of $n(t) \propto r(t)^{-3}$ as a function of 
the radius $r(t)$ in the case of homogeneous isotropic expansion, while it 
scales as $n(t) \propto r(t)^{-D}$ for a fractal volume with Hausdorff 
dimension $D < 3$. The existence of a spatio-temporal evolution, such as
$n(t) \propto r(t)^{-3}$ for adiabatic expansion, implies then also
the prediction of a statistical correlation or scaling law 
$n \propto r^{-3}$, if the expansion speeds $v = \partial r / 
\partial t$ of the sample have a limited range.

Needless to say, that scaling laws obtained from solar data, where we have 
ample spatial resolution, are extremely useful for the interpretation of 
stellar data, where we have no spatial resolution at all and have to rely on 
scaling laws measured in solar or magnetospheric plasmas. Here we focus 
on the spatio-temporal evolution of solar flares, which is a completely 
unexplored topic, but bears important information on the underlying 
dynamic processes. There are very few statistical measurements of 
spatial scales $L$, areas $A$, and volumes $V$, or fractal dimensions $D$ 
of solar flares, and virtually no statistical studies about the temporal 
evolution of these parameters, such as $L(t), A(t), V(t), D(t)$. 
A few statistical measurements of spatio-temporal parameters of solar flares
(compiled in Aschwanden 1999) 
have been made from the S-054 soft X-ray imager onboard {\sl Skylab}
(Pallavicini et al.~1977), from {\sl Yohkoh} soft X-ray images 
(Kano and Tsuneta 1995; Porter and Klimchuk 1995; Aschwanden et al.~1996;
Metcalf and Fisher 1996; Reale et al.~1997; Shimizu 1997; Garcia 1998;
Nagashima and Yokoyama 2006), 
from the {\sl Multispectral Solar Telescope Array (MTSA)} rocket flight 
(Kankelborg et al.~1997), from the extreme-ultraviolet (EUV) imager 
SOHO/EIT (Berghmans et al.~1998; Krucker and Benz 2000), and from the 
EUV imager on TRACE (Aschwanden et al.~2000; Aschwanden and Parnell 2002; 
Aschwanden and Aschwanden 2008a,b). Most of these studies provide statistics
on spatial length scales $L$, areas $A$, and durations $T$ of flares
(down to nanoflares), but there exists no study to our knowledge that
provides statistics on the spatio-temporal evolution $L(t)$ or $A(t)$
of solar flares.

In this paper we are going to analyze the spatio-temporal evolution of all 
large (GOES X- and M-class) flares observed during the first two years of 
the SDO mission, which were observed with
high spatial resolution ($0.6\arcsec$), high cadence (12 s), and in 7 coronal 
wavelengths filters that cover a wide temperature range ($T \approx 
0.5 - 16$ MK). This is an ideal data set for such a study, because 
both spatial and temporal parameters can be measured with unprecedented 
accuracy and 100\% time coverage. Setting a threshold of $>$M1.0 GOES class, 
we obtain a complete set of 155 flare events detected with both GOES and AIA,
which makes it to a perfect representative statistical sample. 
The data analysis presented here is restricted to the 335 \ang\ filter, 
which appears to be very suitable to capture the high-temperature 
component of the energy release and heating phase of these 
largest flares. We fit then various theoretical models to
the measured time profiles of the flare areas 
$A(t)$ or mean radius $r(t) \propto A(t)^{1/2}$, such as:  
classical diffusion, $r(t) \propto t^{1/2}$; sub-diffusion 
$r(t) \propto t^{\beta/2}$ with $\beta < 1$; super-diffusion or 
L\'evy flights, $r(t) \propto t^{\beta/2}$ with $\beta > 1$; or logistic 
growth, where the time profile initially expands exponentially and then 
saturates at a finite level, a limit that is also called {\sl carrying 
capacity} of limited resources in ecological models. The generic form
of these spatio-temporal evolution functions $r(t)$ are shown in Fig.~1.
From the fits of the theoretical models to the observed flare data we aim 
then to gain physical insights into the underlying dynamic processes, 
which concern the spatial propagation (or chain reaction) of 
magnetic reconnection or other nonlinear energy dissipation processes. 
A particular statistical model that we test is the so-called 
{\sl fractal-diffusion self-organized criticality model}, which predicts 
specific spatio-temporal evolutions and powerlaw distributions of the 
spatial and temporal parameters.

The plan of this paper consists of a brief description of relevant 
theoretical models (Section 2), the statistical data analysis and 
forward-fitting to 155 flare events observed with AIA/SDO (Section 3), 
a discussion of the interpretation and consequences of the results 
(Section 4), conclusions (Section 5), and a generalization of the
FD-SOC model (Appendix A).

\clearpage

\section{		THEORETICAL MODELS			}

In the following we introduce some general theoretical models 
that describe the spatio-temporal evolution of physical systems, which
have some universal validity for the dynamics of a large number of 
nonlinear and complex systems. The reason for their universal validity 
and applicability lies in the fact that they are formulated only in terms 
of the fundamental parameters of space and time, while individual 
applications generally involve more specific physical parameters 
(such as densities, temperatures, or magnetic fields in astrophysical 
applications). 

\subsection{Classical Diffusion and Brownian Motion}

Classical diffusion obeys the following differential equation,
\begin{equation}
	{\partial f(x,t) \over \partial t}
	= \kappa {\partial^2 f(x,t) \over \partial x^2} \ ,
\end{equation}
where $f(x,t)$ is the spatio-temporal distribution function of particles
and $\kappa$ is the diffusion coefficient. The diffusion equation, 
expressed here for a 1-dimensional (1-D) space coordinate $x$, can be
generalized to 2-D or 3-D space, where $x$ then signifies the distance $r$
from the center position of the initial distribution. The solution of
the diffusion equation is a Gaussian distribution function (Einstein 1905),
\begin{equation}
	f(x,t)={1 \over (4\pi \kappa t)^{1/2}} \exp^{(-x^2/4\kappa t)} \ ,
\end{equation}
with a time-dependent evolution of its second moment,
\begin{equation}
	\left< x^2 \right> = 2 \kappa t \ ,
\end{equation}
which simply means that its radial size $r(t)=\sqrt{\left< x^2 \right>}$ 
evolves proportionally to the square root of the time,
\begin{equation}
	 r(t) \propto t^{1/2} \ ,
\end{equation}
since the start of the process from an initial $\delta$-function.
This spatio-temporal relationship has originally been applied to the
molecular motion in a fluid (Brown 1828), and thus is also called
{\sl Brownian motion}, named after the Scottish botanist Robert Brown.
Experimentally it can be demonstrated by watching the vibrations of a
dust particle suspended in a fluid under a microscope. While the
random walk of a single particle is stochastic and unpredictable at
every time step, the average distance of the motion of an ensemble of
particles or the radial expansion of the ensemble is well-characterized
by the square-root relationship of Eq.~(4), and is also called classical
diffusion, or more popularly ``the random walk of a drunkard''.
Numerous applications of classical diffusion can be found in nature,
laboratory physics, and astrophysics, such as the 
expansion of released gases, aerosols in the atmosphere, dust clouds from
vulcanos, plasma diffusion in fusion reactors, thermal diffusion in
solar flare plasmas, or insect ecology in biophysics. For a recent 
textbook on applications of Brownian motion see, e.g., 
Earnshaw and Riley (2011).

\subsection{Anomalous Diffusion, Sub-Diffusion, Super-Diffusion, L\'evy Flights}

Deviations from classical diffusion processes, also called {\sl anomalous
diffusion}, are often defined in terms of a non-linear dependence on
time, characterized with a power law index $\beta$ deviating from unity,
\begin{equation}
	r(t) \propto t^{\beta/2} \qquad 
	\left\{ 
	\begin{array}{ll}
		\beta < 1 & \mbox{(sub-diffusion)} \\
		\beta = 1 & \mbox{(classical diffusion)} \\
		\beta > 1 & \mbox{(super-diffusion or L\'evy flight)} 
	\end{array}
	\right.
\end{equation}
We show the generic time evolution of a sub-diffusion process with
$\beta=1/2$ and a super-diffusion process with $\beta=3/2$ in Fig.~1.
Anomalous diffusion implies more complex properties of the diffusive
medium than a homogeneous structure, which may include an 
inhomogeneous fluid or fractal properties of the diffusive medium.
Anomalous diffusion was found in biology, such as active cellular 
transport, protein diffusion within cells, or diffusion through porous 
media (percolation phenomena). 

Related to super-diffusion processes are {\sl L\'evy flights},
named after the French mathematician Paul Pierre L\'evy by
Mandelbrot (1982), which are defined similarly to super-diffusion 
processes (with $\beta > 1$ in Eq.~5) and lead to heavy-tailed 
(powerlaw) distribution functions $f(x,t)$, in excess of the 
Gaussian distribution functions (Eq.~2) obtained for classical diffusion.
A popular example for L\'evy flights is the random walk of sharks or 
other ocean predators, which abandon Brownian motion by occasional 
large-range jumps, when they do not find sufficient food.
Other applications of L\'evy walks involve diffusion on fractal structures,
tracer diffusion in living polymers, turbulent rotating flows, 
electromagnetically driven flows, subrecoil laser cooling, chaos 
in a Josephson junction (electronic chip), or even the geographic travel
of bank notes. For an overview see, e.g., Zumofen et al.~(1999). 

\subsection{Fractal-Diffusion in Self-Organized Criticality Systems}

Inhomogeneous media can be structured in complex patterns. One of the
simplest complex patterns are fractal structures (Mandelbrot 1982),
which can be characterized by a single parameter, such as the
fractal dimension $D$. For instance, a cubic volume $V$ with length $L$
that has a space-filling Euclidean volume $V = L^3$, can have a
fractal substructure with a fractal volume $V = L^D$, with $D<3$
being the (fractal) {\sl Hausdorff dimension}.

Such fractal volumes were found to describe well the avalanches in
nonlinear dissipative systems in the state of self-organized criticality 
(SOC), which led to the {\sl statistical fractal-diffusive model of a 
slowly-driven self-organized criticality system} (Aschwanden 2012).
The prototype of a SOC model is the cellular automaton model, which
numerically simulates the spatio-temporal evolution of a SOC avalanche
by thresholded next-neighbor interactions. The ``spatial diffusion''
of a SOC avalanche is defined by an iterative mathematical re-distribution
rule between next neighbors in a grid $(i,j,k)$, such as,
\begin{equation}
        \begin{array}{ll}
        z(i,j,k)=z(i,j,k)+1 & {\rm initial\ input} \\
        z(i,j,k)=z(i,j,k)-8 & {\rm if}\ z(i,j,k) \ge 8 , \\
        z(i\pm 1,j\pm 1,k\pm 1)=z(i\pm 1,j\pm 1,k\pm 1)+1 &
        \end{array}
\end{equation}
where $z$ denotes the number of energy quanta in each node that are
re-distributed in the nonlinear energy dissipation process of a SOC
avalanche (Bak et al.~1987). It was found that the instantaneous volume 
change $dV(t)/dt$ of a SOC avalanche can be approximated with a fractal 
scaling (Aschwanden 2012),
\begin{equation}
	{dV(t) \over dt} \propto x(t)^{D} \ ,
\end{equation}
and evolves in space according to the classical diffusion law (Eq.~4)
\begin{equation}
	r(t) \propto t^{1/2} \ .
\end{equation}
This scaling law has been inferred from the spatio-temporal evolution of 
a few cellular automaton avalanches (see Fig.~3 in Aschwanden 2012),
which constitutes a strong prediction for any observation of SOC 
systems in nature. This prediction, however, has never been tested with 
observational data, such as for solar flares, which we set out to test 
in this study for the first time.

\subsection{Logistic Growth Model}

Instabilities represent a loss of equilibrium and show often a 
time evolution that consists of an initial exponential
growth phase (which we might call the ``acceleration phase''), followed 
by quenching and saturation of the instability in the decay phase
(which we might call ``deceleration phase''). The total dissipated
energy in such an instability is often approximately proportional to 
the unstable volume. We can then describe the exponential growth phase 
with a proportionality of the change of volume $dV(t)/dt$ to the
instantaneous volume $V(t)$,
\begin{equation}
	{dV(t) \over dt} = \Gamma V(t) = {1 \over \tau_G} V(t)\ ,
\end{equation}
where $\Gamma$ is the growth rate, and $\tau_G=1/\Gamma$ denotes 
the e-folding growth time, which has the simple
solution of an exponential function, 
\begin{equation}
	V(t) = V_0 \exp{({t \over \tau_G})} \ .
\end{equation}
A more general approach to describe both the exponential growth phase
together with the saturation phase is the {\sl logistic equation},
which has been widely used in ecologic applications, but has universal 
validity for nonlinear systems with limited free energy. The logistic 
equation is defined by a simple first-order differential equation, 
discovered by Pierre Fran\c{c}ois Verhulst in 1845 (see textbooks on 
nonlinear dynamics, e.g., May 1974; Beltrami 1987, p.61; Jackson 1989, 
p.75; Aschwanden 2011, p.94),           
\begin{equation}
	{dV(t) \over dt} = {V(t) \over \tau_G} 
	\left[ 1 - {V(t) \over V_\infty} \right] \ ,
\end{equation}
where $V_{\infty}$ represents the asymptotic limit that is reached
in the saturation phase. In ecologic applications, the asymptotic
limit $V_\infty$ is also called {\sl carrying capacity}, such as
the total amount of available energy or ressources that can be sustained
world-wide. From Eq.~(11) we see that the logistic equation
approaches the exponential growth equation (Eq.~9) for small times
(when $V(t) \ll V_\infty$), and the limit $dV(t)/dt \approx 0$ and
$V(t) \approx V_{\infty}$ for large times. The explicit solution of
Eq.~(11) is, 
\begin{equation}
        V(t) = {V_\infty \over 1 + \exp(-{t - t_1 \over {\tau}_G})} \ ,
\end{equation}
while the time derivative $dV/dt$ (representing the energy dissipation
rate $dE/dt$ if the energy is proportional to the volume), 
\begin{equation}
        {dV(t) \over dt} =
                {V_\infty \over {\tau}_G} {\exp(-{t - t_1 \over {\tau}_G})
                \over [ 1 + \exp(-{t - t_1 \over {\tau}_G}) ]^2 } \ .
\end{equation}
If we approximate the unstable volume with a spherical geometry, the
spatial length scale or radius $r(t)$ is related to the volume $V(t)$ as,
\begin{equation}
	V(t) = {4 \over 3}\pi r(t)^3 \ , \qquad
	V_\infty = {4 \over 3}\pi r_{\infty}^3 \ ,
\end{equation}
which predicts the following spatio-temporal evolution $r(t)$,
\begin{equation}
	r (t) = {3 \over 4 \pi} V(t)^{1/3} =
        r_{\infty} \left[ 1 + \exp(-{t - t_1 \over {\tau}_G}) \right]^{-1/3} \ .
\end{equation}
A graphical illustration of this logistic growth function is shown in Fig.~1,
which appears to conform to a lower limit of sub-diffusion processes
(for $\beta \mapsto 0$). The logistic growth model was applied to
population growth, neural networks, tumor growth in medicine, autocatalytic
reactions in chemistry, the statistical distribution of fermions in
atomic physics, language change in social sciences, or diffusion of
innovations in economics. In solar physics, the logistic-growth model
was applied to model hard X-ray pulses (Aschwanden et al.~1998) and
magnetic energy storage in active regions (Wang et al.~2009).

\subsection{Combining Logistic Growth and Diffusion Models}

In order to have compatible models for the exponential growth phase,
we adopt the same exponential growth phase of the logistic model (Eq.~15)
for the diffusion models, so that the spatio-temporal evolution $r(t)$
differs only in the deceleration phase for the various models shown
in Fig.~1. We choose the transition time between the acceleration
and deceleration phase at $t=t_1$ in the formulation of the logistic
growth curve (Eq.~15), which has the value $r_1$ and time derivative $v_1$,
\begin{equation}
	r_1 = r(t=t_1) = 2^{-1/3} r_{\infty} \ ,
\end{equation}
\begin{equation}
	v_1 = {dr \over dt} (t=t_1) 
	= {2^{-4/3} \over 3} {r_{\infty} \over \tau_G} 
	= {r_1 \over 6 \tau_G} \ .
\end{equation}
Defining the diffusion radius $r(t)$ in the deceleration phase with
a diffusion constant $\kappa$,
\begin{equation}
	r(t) = \kappa \ (t-t_k)^{\beta/2} \ ,
\end{equation}
which has the following value $r_1$ and time derivative $v_1$ at the
transition time $t=t_1$,
\begin{equation}
	r_1 = r(t=t_1) = \kappa (t_1-t_k)^{\beta/2} \ ,
\end{equation}
\begin{equation}
	v_1 = {dr \over dt} (t=t_1) 
	= {\kappa \beta \over 2} \left({r_1 \over \kappa}\right)^{1-2/\beta} \ ,
\end{equation}
which have to match the boundary conditions $r_1$ and $v_1$ (Eqs. 16-17)
of the exponential growth phase. From Eqs.~(16)-(20) we obtain the
constants $t_1, r_\infty, \tau_G$,
\begin{equation}
	t_1 = t_k + \left({r_1 \over \kappa}\right)^{2/\beta} \ ,
\end{equation}
\begin{equation}
	r_\infty = r_1 \ 2^{1/3} \ ,
\end{equation}
\begin{equation}
	\tau_G = {r_1 \over 6 \ v_1} \ .
\end{equation}
Thus the combined time profile $r(t)$ with a smooth transition from the
exponential growth (acceleration) phase to the diffusion (deceleration)
phase is,
\begin{equation}
	r(t) = \quad
	\left\{ 
	\begin{array}{ll}
        r_{\infty} \left[ 1 + \exp(-{t - t_1 \over {\tau}_G}) \right]^{-1/3} 
	& {\rm for} \ t \le t_1 \\
	\kappa (t-t_k)^{\beta/2} 
	& {\rm for} \ t > t_1 
	\end{array}
	\right.
\end{equation}
which has four free parameters for fitting, where we treat 
$(\kappa, t_k, \beta, r_1)$ as independent paramters, while
$(t_1, r_\infty, \tau_G)$ are constrained by the boundary conditions 
(Eqs.~21-23).  For classical diffusion ($\beta=1$) the number of free 
parameters reduces to three.

\section{OBSERVATIONS AND DATA ANALYSIS}

\subsection{GOES and AIA Observations}

We select all solar flare events detected with the {\sl Geostationary 
Operational Environmental Satellites (GOES)} and the {\sl Atmospheric
Imaging Assembly (AIA)} on the {\sl Solar Dynamics Observatory (SDO)}
(Lemen et al.~2012)
above a threshold of the M1.0 class level (which includes M- and X-class 
events) during the first two years of the SDO mission. The selected
time era starts when the first science data from AIA became available,
13 May 2010, and ends on 31 March 2011 when we started the data analysis.
The selected time period contains a total of 155 solar flare events
larger than M1.0 class, including 12 events larger than X1.0 class.
The flare time intervals, listed in Table 1, were taken from the official
GOES flare catalog. We extracted AIA images in all 6 coronal wavelengths
during the GOES flare time intervals, with AIA data being available
in 100\% of the cases, but restricted our analysis to the 335 \ang\ filter 
(Fe XVI), which appears to be the most suitable filter to probe the
high-temperature component of these largest flares, since the other
filters that are sensitive to high temperatures (94 \ang\ and 131 \ang)
have a double response to cooler plasma of $T\lapprox 1.0$ MK that is
not yet well-calibrated (due to the incompleteness of the CHIANTI atomic 
data base for Fe lines at temperatures $\lapprox 10^6$ K).
All AIA images have a cadence of $\Delta t=12$ s and a pixel size of 
$\Delta x=0.6\arcsec \approx 435$ km, which corresponds to a spatial
resolution of $2.5 \Delta x = 1.5\arcsec \approx 1100$ km. 
We normalized all AIA 335 \ang\ images by the exposure time.
The total number of analyzed images amounts to 11,767 AIA 335 \ang\
images, which averages to 76 images per flare event and an average
flare duration of 910 s.

\subsection{Data Analysis Method}

In order to quantify the spatio-temporal evolution of flares we measure
the flare area above some threshold flux level, which was chosen in the
AIA 335 \ang\ filter at a constant level of $F_{thresh}=100$ and 200 DN/s. 
This is well below the median flux value of $F_{med}(t=0)=518$ DN/s and 
mean of $<F(t=0)>=687 \pm 645$ DN/s of the maximum flux values of the 
EUV images at the flare start time, defined by GOES. Thus the instantaneous
flare area $a(t)$ at time $t$ is defined by the number of pixels that
have a flux $F$ in excess of the threshold $F_{thresh}$, 
\begin{equation}
	a(t_i) = N[F_{x,y}(t_i) \ge F_{thresh}] \ ,
\end{equation}
while the time-integrated flare area $A(t)$ is the combined area of all
spatially overlapping instantaneous flare areas $a(<t)$,
\begin{equation}
	A(t_i) = a(t=0) \oplus a(t=1) \oplus ... \oplus a(t=t_i) \ ,
\end{equation}
where the symbol $\oplus$ indicates a logical OR-function between
the pixels contained in each instantaneous flare area before time $t_i$.
Thus, the time-integrated flare area $A(t)$ is a monotonically increasing
quantity that contains all flux pixels that exceeded the threshold 
$F_{thresh}$
at any time before a given time $t_i$, i.e., during the flare time
interval $0 \le t \le t_i$. 

The spatial scale $r(t)$ of each (time-integrated) flare area $A(t)$ is
then defined by the radius of an equivalent circular area,
\begin{equation}
	r(t) = \sqrt{ A(t)/ \pi } \ .
\end{equation}
The advantage to use the time-integrated flare area $A(t)$ over the 
instantaneous flare area $a(t)$ is the robustness against temperature
effects such as conductive or radiative cooling, which can decrease the
flux of flaring pixels below the flux threshold after some time.
However, we tested our method for both options and found compatible
results in the cases of flares with no rapid cooling.

\subsection{Examples of Analyzed Flares}

In Fig.~2 we show an example of the data analysis of a single event
($\#28$), out of the 155 analyzed events. The GOES time profiles are
shown on a linear flux scale (Fig.~2, top panel), with a start time of
2011 March 07, 19:43 UT, peak time 20:12 UT, end time 20:58 UT, and
GOES class M3.7 ($=3.7 \times 10^{-5}$ W m$^{-2}$), according to the 
official NOAA flare catalog. The flare duration is over an hour ($T=4500$ s).

At this point it might be appropriate to recall the definition of
the flare duration used in the NOAA flare detection algorithm: The event 
starts when 4 consecutive 1-minute X-ray values have met
all three of the following conditions: (i) All 4 values are above the B1
threshold; (ii) All 4 values are strictly increasing; (iii) The last value
is greater than 1.4 times the value that occurred 3 minutes earlier.
The peak time is when the flux value reaches the next local maximum.
The event ends when the current flux reading returns to half of the peak
value (http://www.ngdc.noaa.gov/stp/solar/solarflares.html).

We analyzed the same flare time interval from AIA 335 \ang\ data, 
which amounts, with a cadence of $\Delta T=12$ s, to a total of
$N_T=T/\Delta T =4500/12=375$ images.
The total solar flux $F_{sun}(t)$ during this time interval (Fig.~2,
second panel) is shown along with the flux $F_{flare}(t)$ integrated
over the flare area (approximately corresponding to the field-of-view
shown in the bottom rows of Fig.~2). The agreement between the two
flux profiles confirms the correct localization of the flare position,
which is centered at the heliographic position N30/W48 for this event.

The flare area $A(t)$ is evaluated by counting the pixels above the
threshold $F_{thresh}=100$ DN/s, (or 200 DN/s, respectivtly). The pixels
of the time-integrated area combines all overlapping
flare areas from earlier times $t>t_0$, since flare start. 
The radius $r(t)=\sqrt{A(t)/\pi}$ of the
equivalent circular (monotonically growing) flare area $A(t)$ is shown
in the third panel of Fig.~2 (histogrammed curve), fitted with the
model $r(t)$ defined in Eq.~(24) (shown with a solid curve in Fig.~2). 
The goodness-of-fit is simply 
evaluated from the average deviation normalized by the maximum flare 
radius, $r_{max} = max[r(t)] = r(t_{end})$, 
\begin{equation}
	\Delta r/r_{max}= {< r_{obs}(t) - r_{model}(t) > \over r_{max}} \ ,
\end{equation}
which is $\Delta r/r_{max}=1.3\%$ for the threshold of $F_{thresh}=100$
DN/s, or $\Delta r/r_{max}=1.2\%$ for the threshold of $F_{thresh}=200$
DN/s, respectively. We determine the flare area for two different
thresholds, in order to test the uncertainty and sensitivity of the
best-fit model parameters on the threshold value.
In this flare we obtain a diffusion coefficient of $\kappa=117\pm10$
pixel s$^{-1/2}$ and a diffusion index of $\beta=0.47\pm0.08$, which
demonstrates a relatively small uncertainty caused by the chosen two 
fixed flux thresholds (see values given in Fig.~2 and listed in Table 1).

Five snapshots of the time-integrated flare images are shown in the bottom 
panels of Fig.~2, rendered as preflare-subtracted flux (Fig.~2, fourth row), 
as well as highpass-filtered flux (Fig.~2, bottom row), showing the
flare loop fine structure with enhanced contrast.
The contours of the flare area $A(t)$ are
also shown as black contour lines in the flare snapshot images. The
flux $A(t=0)$ is near zero at the beginning of the time series, indicating that 
the flare start time (defined by GOES) corresponds to an early phase 
when the 335 \ang\ flux was near the chosen threshold.

The data analysis is also shown in condensed form for a selection of 
another 48 events (out of the 155 analyzed events) in Fig.~3, where we
show the spatio-temporal evolution of the GOES 1-8 \ang\ flux 
$f_{GOES}(t)$ (dotted curve), the EUV flux $f_{AIA}(t)$ of the 335 \ang\
filter (dashed curve), and the spatial evolution of the radius $r(t)$ of
the equivalent circular flare area (histograms in Fig.~3), along with
the best fit of the diffusion model (Eq.~24) for the two flux thresholds
of $F_{thresh}=100$ and 200 DN/s (solid curves in Fig.~3), as well as a
(preflare-subtracted) time-integrated AIA 335 image at the end of the 
flare (right panels in Fig.~3). The examples show the enormous variety
of morphological shapes of flare areas, which generally consist of
groups or arcades of post flare loops (in different projections) with
highly fractal substructures. We show also the contours of the final 
time-integrated flare area above a threshold of $F_{thresh}=100$ DN/s
in Fig.~3 (right panels), which generally encompasses a contiguous
flare area. The time profiles reveal that the GOES flux generally
peaks earlier than the EUV flux, although they start at similar times.
The evolution of the flare radius $r(t)$ shows often a first expansion
triggered by a precursor flare. In such cases, our fits apply to the
main phase, while the area associated with the precursor is subtracted.
Note the variety of $\beta$-values in different flares, ranging from
the logistic limit $\beta \gapprox 1$ to the classical diffusion limit
of $\beta \approx 1$. The fits of the diffusion model $r(t)$ shown
in Fig.~3 illustrate the adequacy of the functional form (Eq.~24) that
fits the data always within a few percent accuracy, as well as the
consistency of results obtained from different thresholds.

\subsection{Statistical Results}

The statistics of the observed parameters is visualized in form of
histogram distribution functions shown in Fig.~4, as well as in form
of correlations between pairs of parameters in Fig.~5. The best-fit
parameters $L$, $\kappa$, $\beta$, and $q_{fit}=\Delta r/r_{max}$
are also listed in Table 1 for each flare, and the ranges, means,
standard deviations, and medians are summarized in Table 2. 

The length scale $L$, defined as the spatial radius of the flare area
after subtraction of possible preflare areas, 
i.e., $L = \sqrt{r(t_{end})^2-r(t_{start})^2}$, was found in a range
of $L=5-50$ Mm. The flare time scale $T$, which we define from the 
GOES flare start time to the peak time, e.g., $T=t_{peak}=t_{start}$,
has a range of $T=120-8460$ s (i.e., from 2 min to over 2 hours). 
The GOES flux ranges for this sample of M and X-class flares from
$F_{GOES}=10^{-5}$ to $6.9 \times 10^{-4}$ W m$^{-2}$.
The AIA 335 \ang\ flux is found in the range of $F_{AIA}=7 \times 10^4$
to $2.5 \times 10^7$ DN/s. Then we measured also the average expansion
velocity $v=r_{max}/(t_{start}-t_{end})=3.3-103$ km s$^{-1}$, as well
as the maximum velocity during the flare time interval,
$v_{max}=8-550$ km s$^{-1}$. These parameters $(L, T, F_{GOES},
F_{AIA}, v, v_{max}$) have approximate powerlaw distributions 
(Fig.~4, left and middle column).  

The best-fit model parameters have nearly Gaussian distributions
(Fig.~4, right column), which can be characterized by the mean and 
standard deviations: the diffusion coefficient $\kappa=56 \pm 24$ 
km s$^{-\beta/2}$, the diffusion index $\beta=0.53 \pm 0.27$, and the 
goodness-of-fit $\Delta r/r_m=2.2 \pm 0.7\%$. Thus, the diffusion
index, which is found in a range of $\beta=0.04-1.35$, covers the
entire range from logistic growth ($\beta \approx 0$), to sub-diffusion
($\beta \approx 1/2$), and classical diffusion ($\beta \approx 1$), 
but extends much rarer into the regime of super-diffusion or L\'evy flights 
($\beta \approx 3/2$). No case was found that approaches the limit 
of linear expansion ($\beta = 2$).

\subsection{Parameter Correlations}

Parameter correlations can reveal physical scaling laws. In Fig.~5
we show scatterplots between pairs of observed parameters 
($L, V, T, F_{GOES}, F_{AIA}, v, v_{max}, \kappa, \beta$).
Let us first discuss how these parameters depend on the fundamental
length scale $L$ of flares (Fig.~5).
The flare duration $T$ shows only a loose correlation 
with the length scale $L$, i.e., $T \propto L^{2.0\pm1.2}$ (panel a), 
or $L \propto T^{0.8\pm 0.5}$ (panel d). 
Weak correlations are also found for the GOES flux $F_{GOES}(L)$ (panel b), 
the mean expansion velocity $v(L)$ (panel c), as a function of
the length scale $L$. The strongest correlations are found between
the AIA 335 \ang\ flux and the length scale, i.e., $F_{AIA}(L) \propto
L^{2.4\pm0.5}$ (panel e), or with the volume $F_{AIA}(V) \propto 
V^{0.8\pm0.2}$ (panel g), and the diffusion coefficient 
$\kappa(L) \propto L^{0.9\pm0.1}$ (panel f). The former good correlation can be
explained if the EUV emission is proportional to the total flare volume,
in which case we expect $F_{AIA} \propto V \propto L^3$. The latter
good correlation can be explained for sub-diffusion and logistic growth,
where the diffusion equation $L \propto \kappa (t-t_1)^{\beta/2}$
shows only a weak dependence on the time duration due to the small value
of the power index $\beta/2 \ll 1$, which yields almost a proportionality
of $L \propto \kappa$. 

Correlating all other parameter combinations we found five more cases
that are worthwhile to mention, shown in the lower half of Fig.~5.
The mean expansion velocity scales almost reciprocally with the flare
duration, $v_{max}(T) \propto T^{-1.0\pm0.6}$ (panel h), which is consistent
with a maximum speed that is independent of the length scale, so that
the mean length scale $<L>$ is a constant (with uncorrelated scatter) and 
yields $v_{max}=<L>/T \propto T^{-1}$. The maximum velocity shows also a good
correlation with the mean velocity, i.e., $v_{max}(v) \propto v^{1.4\pm0.4}$
(panel k). There is also an expected trend of correlated EUV and soft X-ray 
fluxes, i.e., $F_{AIA} \propto F_{GOES}^{1.9\pm1.4}$ (panel i). What is most 
interesting that comes out of this study, is that the diffusion coefficient 
is strongly correlated with the EUV flux, i.e., $\kappa \propto 
F_{AIA}^{0.4\pm0.1}$ (panel l), as well as with the soft X-ray flux, i.e., 
$\kappa \propto F_{GOES}^{0.6\pm0.3}$ (panel j). 

\section{DISCUSSION}

In this study we measured spatial $(L)$ and temporal scales $(T)$ in 
solar flares and investigated for the first time their spatio-temporal 
evolution $L(t)$. We quantified the spatio-temporal evolution with a
general diffusion equation that is quantified in terms of a diffusion
coefficient $\kappa$ and a diffusion powerlaw index $\beta$ (Eq.~24).
The theoretical range of the diffusion powerlaw index, $\beta=0-2$
includes the limit of logistic growth $(\beta \approx 0)$, sub-diffusion
($\beta \approx 1/2$), classical diffusion ($\beta=1$), and 
super-diffusion or L\'evy flights ($\beta=3/2$). In the following we
discuss the consequences of the results in the context of
self-organized criticality models (section 4.1), diffusion processes
in the photosphere (section 4.2), in the corona (section (4.3), and 
in solar flares (section 4.4).

\subsection{Self-Organized Criticality Models}

The size distribution of solar flare parameters (energy, peak count rates,
durations) follow all powerlaw distribution functions that have been
interpreted in terms of a slowly-driven nonlinear system in the state
of self-organized criticality (SOC) (Lu and Hamilton 1993). A quantitative
model that predicts the values of the powerlaw slopes and the associated 
correlations of physical parameters has been recently proposed, was tested 
with numerical simulations of cellular automaton models, and compared with 
solar flare statistics (Aschwanden 2012).
The most fundamental parameters of SOC models are the spatial size (e.g., 
length scale $L$, area $A$, volume $V$) and the time duration $T$.
In SOC models applied to astrophysical observations, generally a
proportionality is assumed between the spatial volume $V$ and the 
radiated fluence $E$, which can be justified by the linear relationship 
between the column depth $\int dz$ and emission measure 
$EM \propto \int n_e^2 dz$ for optically-thin EUV and soft 
X-ray emission. Consequently, the instantaneous fractal volume of a SOC
avalanche $(dV_S/dt)$ corresponds to the energy dissipation rate $(dE/dt)$ 
or flux $F$, and the peak energy dissipation rate $(dE/dt)_{max}$ to 
the peak flux $P$.

While classical diffusion ($L \propto T^{1/2}$) was adopted in the 
{\sl statistical fractal-diffusive avalanche model of a slowly-driven 
self-organized criticality system (FD-SOC)} (Aschwanden 2012), based 
on empirical evidence from cellular automaton models, we generalize
the model for anomalous diffusion and for wavelength-dependent
flux-volume scaling law $F_{\lambda} \propto (dV_S/dt)^\gamma$ 
in Appendix A, which predicts slightly
different correlation coefficients and powerlaw slopes of the distribution
functions. In 3D-space ($S=3$), for which a mean fractal dimension of
$D_3=2.0$ is predicted, and using the measured range of the diffusion
index $\beta \approx 0.1-1.0$ (Fig.~4) and $\gamma=0.8$ (Fig.~5, panel g), 
the following powerlaw slopes are predicted
by the generalized FD-SOC model (Appendix Eq. A14),
\begin{equation}
        \begin{array}{ll}
        \alpha_L &=  3.0                               \\
        \alpha_T &=  1+(S-1) \beta/2  \approx 1.1-2.0  \\
        \alpha_F &=  1+(S-1)/D_S \gamma = 2.25	       \\
        \alpha_P &=  1+(S-1)/S\gamma    = 1.83         \\
        \alpha_E &=  1+(S-1)/(D_S \gamma +2/\beta) \approx 1.1-1.6 \\
        \end{array} \ .
\end{equation}
For the size distributions of flare length scales $L$ we expect
in the framework of the FD-SOC model $\alpha_L=3$, while our observations
show $\alpha_L=2.0$ (Fig.~4, first row, left panel).
For the size distribution of flare durations $T$, classical
diffusion predicts a slope of $\alpha_T=2.0$ and sub-diffusion (with
$\beta=1/2$) predicts $\alpha_T=1.5$, while our observations show
$\alpha_T \approx 2.2$ (Fig.~4, second row, left panel). 
For the GOES flux $F_G$ and AIA 335 flux $F_A$ we measure powerlaw slopes
of $\alpha_F=1.92$ and $\alpha_F=1.34$ (Fig.~4), while the FD-SOC model
predicts $\alpha_F=2.0$. In a much larger sample of over 300,000 GOES
flares, a value of $\alpha_F=1.98\pm0.11$ was found (Aschwanden and 
Freeland 2012), which is fully consistent with the FD-SOC model
for classical diffusion ($\beta=1$).
Considering the small statistical sample of 155 flares analyzed here,
the obtained powerlaw slope values seem to be not inconsistent 
with the FD-SOC model, given the relatively small range of less than 
a decade over which a powerlaw slope could be fitted.

Alternatively, we can compare the observed parameter correlations (Fig.~5)
and compare with the predictions of the SD-SOC model (Appendix A).
The most fundamental relationship for anomalous diffusion predicts
$L \propto T^{\beta/2}$, whith amounts to $L \propto T^{1/4}$ for
sub-diffusion ($\beta=1/2$), while our linear regression fit yields
$L \propto T^{0.8\pm0.5}$ (Fig.~5, panel d), which has a large uncertainy 
due to the weak dependence and large scatter of $L$ and $T$ values.
A tighter correlation is found between the EUV flux and the flare volume,
i.e., $F_{AIA} \propto V^{0.8\pm0.2}$, which corroborates the standard
assumption in SOC models that the observed flux or emission is approximately 
proportional to the emitting volume. 

In summary, the small statistical sample of 155 flare events does not allow
us to determine the statistical scaling laws (or correlations) between 
spatial $L$ and temporal scales $T$ or the powerlaw distributions 
$N(L)$ and $N(T)$ of spatial and temporal parameters with sufficient 
accuracy, and to test SOC models based on their statistical distributions.
However, our study can quantify the spatio-temporal evolution $L(t)$ 
for individual flare events very accurately, which could potentially be 
compared with the spatio-temporal evolution $L(t)$ of cellular automaton 
simulations to test SOC models quantitatively. This could clarify whether
SOC cellular automaton avalanches based on (thresholded) next-neighbor 
interactions can be statistically described by anomalous or classical 
diffusion processes. Diffusion-dominated dynamics (Bregman 2007, 2008)
or diffusion entropy (Grigolini et al.~2002) has been applied to SOC
avalanche models elsewhere. 

\subsection{Diffusion Processes in the Photosphere}

Magnetic fluxtubes that emerge through the photosphere and build
footpoints of coronal magnetic loops as well as postflare loops,
are buffeted around by the random walk dynamics of subphotospheric
magneto-convection. Measurements of the random motion of granules
and supergranulation features have been characterized with a diffusion
equation,
\begin{equation}
	<r^2> = 4 D t \ , 
\end{equation}
where the diffusion coefficient $D$ is related to our definition 
$\sqrt{<r^2>}=\kappa t^{1/2}$ (Eq.~18) by 
$D=(\kappa/2)^2$. Thus, our measurements in solar flares
with $\kappa=56 \pm 24$ km s$^{-1/2}$ would correspond to a range of
$D \approx 300-1600$ km$^2$ s$^{-1}$. In comparison, photospheric random
motion has been determined to $D=110$ km s$^{-1}$ in strong-field cores
of active regions, and $D=250$ km s$^{-1}$ in the surrounding area
(Schrijver and Martin 1990), $D=200-400$ km s$^{-1}$ from the motion
of magnetic elements (Mosher 1977), $D=120-230$ km s$^{-1}$ from Quiet
Sun magnetic patterns (Komm et al.~1995), which appears substantially lower 
than in flares. However, higher values have been inferred from the magnetic
polarity reversal during a solar cycle ($D=1100$ km s$^{-1}$;
Leighton 1964, 1969), from meridional flows ($D=600$ km s$^{-1}$;
Wang et al.~1989; Zirin 1985), or from mottles ($D=600$ km s$^{-1}$
corrected for unresolved small-scale elements; Schrijver et al.~1996).
In analogy to our fractal-diffusive
SOC model (Aschwanden 2012), the dispersion of magnetic field elements
across the solar surface was thought to be a diffusion process on a fractal
geometry (Lawrence 1991). The diffusion of magnetic elements across the
solar surface was actually found to have a 2-D fractal dimension of 
$D_2=1.56\pm0.08$ and an anomalous diffusion index of $\beta=0.25\pm0.40$,
similar to our result of preferential sub-diffusion ($\beta=0.53\pm0.27$).
In contrast, Ruzmaikin et al.~(1996) found super-diffusion in photospheric
random motion. A super-diffusive regime with $\beta \approx 1.48-1.67$
was also measured from the proper motion of bright points (Abramenko
et al.~2011). Diffusion in the photosphere is also thought to be
responsible for magnetic flux cancellation (Litvinenko 2011).

In summary, most diffusion processes observed in magnetic features in the
photosphere are generally slower ($D \approx 100-600$ km s$^{-1}$)
than what we observe in flares ($D \approx 300-1600$ km$^2$ s$^{-1}$),
and thus are unlikely to explain the observed spatial diffusion during
flares.

\subsection{Diffusion Processes in the Corona}

The magnetic field evolution in the solar corona is usually described
by the induction equation in the framework of ideal magnetohydrodynamics 
(MHD),
\begin{equation}
                {d {\bf B} \over dt}
                = \nabla \times ({\bf v} \times {\bf B})
                + \eta \nabla^2 {\bf B} \ ,
\end{equation}
with $\eta = {c^2 / 4 \pi \sigma}$ being the magnetic diffusivity. 
The first term on the right-hand side is called the convective term,
while the second term is called the diffusive term. Depending on the
value of the Reynolds number $R_m$, which gives the ratio of the
convective term ($\propto {v}_0 B_0 / l_0$) to the diffusive term
($\propto \eta B_0 / l_0^2$),
\begin{equation}
        R_m = {l_0 {v}_0 \over \eta} \ ,
\end{equation}
the induction equation can be approximated in the two limits by
\begin{equation}
                {d {\bf B} \over dt} \approx
                 \nabla \times ({\bf v} \times {\bf B})
		        \qquad {\rm for}\ R_m \gg 1 \
\end{equation}
\begin{equation}
                {d {\bf B} \over dt} \approx
                 \eta \nabla^2 {\bf B} \
        \qquad {\rm for}\ R_m \ll 1 \
\end{equation}
The plasma in the solar corona is close to a perfectly
conducting medium (with a high Reynolds number $R_m \approx 10^8-10^{12}$),
so that approximation (Eq.~33) with $R_m \gg 1$
applies, while the diffusive limit (Eq.~34) with $R_m \ll 1$ is
not relevant. Thus, cross-field diffusion is generally
inhibited in the solar corona. 

While cross-field transport is generally inhibited 
in low plasma-$\beta$ parameter regions, it can be enabled by gyro-orbit
perturbations due to local magnetic field fluctuations, a process called
Rechester-Rosenbluth diffusion (Galloway et al.~2006), but this process
is estimated to operate on time scales of days, and thus is way too slow 
to explain the observed anomalous diffusion in flares. 
A scenario with small-scale random footpoint motion that injects
energy into the corona by twisting and braiding was invoked as a coronal
heating mechanism, where helicity conservation leads to hyper-diffusion
(VanBallegooijen and Cranmer 2008), but this is also driven by photospheric
diffusion speeds, and thus not fast enough to drive solar flares.

\subsection{Diffusion Processes in Solar Flares}

Let us discuss first diffusion processes in the preflare phase. 
Resistive diffusion of
force-free magnetic fields has been considered to develop infinite field
gradients and to trigger the eruption of solar flares (Low 1973a,b; 1974b)
and acceleration of particles (Low 1974a). On the other side, photospheric
diffusion is thought to reduce the non-potential magnetic (free) energy
that is built up by the photospheric shearing motion (Wu et al.~1992). 
The eruption of a coronal mass ejection (CME) was simulated with MHD
simulations, where slow turbulent diffusion of the footpoints of coronal
magnetic field lines leads to a loss of equilibrium and drives the
eruption (Amari et al.~2003). However, all these diffusion processes
occur before a flare, and thus do not explain the rapid diffusion
during flares.

What happens during flares?  The driving instability of solar flares 
is generally linked to the dynamics of magnetic reconnection processes. 
Diffusion of particles is most efficient in the X-point of collisionless
magnetic reconnection regions, where the magnetic field drops to zero.
This diffusion region is subject of numerous theoretical and simulation
studies (e.g., see review by Hesse et al.~2011). The duration of a
magnetic reconnection process is approximately the Alfv\'enic transit
time across a coronal reconnection region, which is of order
$T_1 = L/v_A \approx 10$ s (for $L=10,000$ km and $v_A=1000$ km s$^{-1}$).
While such a single magnetic reconnection episode (Sweet-Parker-type 
or Petschek-type) can only explain very short single-loop flares,
large flares (with a range of durations $T=120-8460$ s observed here) 
require a chain reaction of magnetic reconnection events, perhaps 
in the order of $N=T/T_1= 12-850$ reconnection episodes. Evidence
for such multi-reconnection events has been demonstrated for numerous
large flares, the most prominent case being the Bastille-Day flare,
where over $N \gapprox 200$ postflare loops have been traced,
probably each one being a remnant of a local magnetic reconnection
process (e.g., Aschwanden and Alexander 2001), forming a multi-loop
postflare configuration (Hori et al.~1998). Theoretical models
of multi-reconnection or intermittent, unsteady, bursty reconnection 
have also been formulated in terms of the tearing mode instability 
and coalescence instability, and combinations of both 
(e.g., Furth et al.~1963; Sturrock 1996; Kliem 1990, 1995; 
Leboef et al.~1982; Tajima et al.~1982; Karpen et al.~1995;
Kliem et al.~2000; Drake et al.~2009). The question is now how the
time evolution of such an intermittent chain reaction of magnetic 
reconnection episodes ties into our simplified diffusion evolutionary 
model. 

Perhaps our measurements of the average expansion speed can give us
a hint about the physical mechanism of the diffusion process. 
We measured average speeds in
the range of $v=3-103$ km s$^{-1}$, and maximum speeds of
$v_{max}=8-550$ km s$^{-1}$. If we assume typical plasma temperatures 
of $T \approx 10-30$ MK for the observed large (M and X-class) flares,
we expect sound speeds of $c_s \approx 166 \sqrt{T_{MK}} \approx 500-900$
km s$^{-1}$, which are about 1-2 orders of magnitude faster than the
observed speeds. Therefore, the diffusive speed we observe in these
large flares cannot just be the propagation speed of upflowing plasma
into and along flare loops initiated by the chromospheric evaporation process, 
which occurs approximately with sound speed. There are some additional 
delays that intervene between the subsequent filling of individual
loops in a flare arcade. A model where subsequent flare loops are
triggered by a slow (acoustic) wave that propagates at some angle
($25^\circ-28^\circ$) along the arcade was proposed by Nakariakov
and Zimovets (2011) and tested by the timing of the hard X-ray
footpoints (Inglis and Dennis 2012), but the predicted hard X-ray
pulse periods did not agree with the observed ones. Nevertheless,
the mean footpoint speed found in the analyzed three flares ($v=5-60$
km s$^{-1}$; Inglis and Dennis 2012) is similar to our observed range 
($v=3-103$ km s$^{-1}$), which
corroborates the congruence between the magnetic reconnection path
and the hard X-ray footpoint path. Similar propagation speeds of the
hard X-ray footpoints were also measured along an arcade of flare loops
by Grigis and Benz (2005), i.e., a mean velocity of $v=63-65$ km s$^{-1}$
parallel to the arcade, with a peak speed of about $v=110$ km s$^{-1}$
for 2 minutes. Hence we can interpret our diffusion speed as mean 
propagation speed of subsequently triggered magnetic reconnection sites.

The measurement of the spatio-temporal evolution during solar flares
should also ameliorate estimates of the {\sl particle number problem} 
(Brown and Melrose 1977), which requires the knowledge of the flare 
volume and its temporal change and spatial propagation path.

\subsection{Fractal Diffusion, Sub-Diffusion, or Logistic Growth ?}

We combined the different anomalous and classical diffusion models
into a single framework (Eq.~24) that has only one single parameter
($\beta$) that discriminates between the different models:
$\beta=0$ is the limit of logistic growth,
$\beta=0-1$ is the sub-diffusive regime,
$\beta=1$ is classical diffusion or Brownian motion,
$\beta=1-2$ is the super-diffusive regime or L\'evy flights,
and $\beta=2$ is the limit of linear expansion.
Our results yielded a range of $\beta=0.04-1.35$ that covers
three of these regimes, but the majority is found in the sub-diffusive
regime with a mean and standard deviation of $\beta=0.53\pm0.27$.
What does this mean and why is it different from the classical
diffusion model based on random walk?

Brownian motion or classical diffusion is based on homogeneous
isotropic expansion. In solar flares, as well as in cellular automaton
models, however, the medium is inhomogeneous (which can be characterized 
by a fractal dimension). The solar corona, moreover, is highly
anisotropic due to the structuring by the magnetic field (in the
low plasma-$\beta$ regime), which also applies to flares, except
to a reduced degree in the diffusion regions of magnetic X-point
reconnection configurations. Let us assume the extreme case where
plasma can only flow in one direction along magnetic fluxtubes.
The anisotropic volume increase will scale as $V(t) \propto r(t)$,
instead of $V(t) \propto r^3$ in an isotropic medium. The observable
area $A(t)$ will then grow only in one direction (say in $x$-direction)
with a constant width (say $\Delta y$),
\begin{equation}
	A_{iso}(t) = x_{iso}(t) \Delta y \ ,
\end{equation}
and the derived flare expansion radius $r(t)=\sqrt{A(t)/\pi}$ will
scale as
\begin{equation}
	r(t) = \sqrt{ {A_{iso}(t) \over \pi} }
		   \propto x_{iso}(t)^{1/2} \propto t^{1/4} \ ,
\end{equation}
for a classical diffusion process in one dimension, i.e., 
$x_{iso}(t) \propto t^{1/2}$. Thus by defining anomalous diffusion
by $r(t) \propto t^{\beta/2}$ we end up with a diffusion index of
$\beta=1/2$, which is indicative of the sub-diffusive regime.
Now, although the electron and ion diffusion region in magnetic
reconnection X-points allow 2-D or 3-D diffusion, our observational
result of sub-diffusion with $\beta \approx 1/2$ suggests that 
the triggering of subsequent magnetic reconnection episodes occurs
along anisotropic 1-D paths and dominates the overall diffusion of 
the flare process, regardless of the near-isotropic diffusion regions
in X-points. 

What does the extreme limit of logistic growth mean, where the
diffusion comes to a halt at some time near the peak time of the
flare and does not expand further. Apparently, the energy release
process approaches regions with strong magnetic fields that are stable 
and not prone to magnetic reconnection, so that the chain reaction of
reconnection episodes does not propagate further, while energy
release in the previous area still continues. It is like
a domino chain reaction that hits the wall of a room. The domino
effect can still propagate in the unstable regions inside the room,
but cannot expand outside of the room. Alternatively, the progressing
flare could also reach the boundary of an active region, where only
weak-field regions of the Quiet Sun are available which harbour
much less non-potential magnetic energy than the inside of 
active regions. 

In summary, the diffusion index $\beta$ tells us some interesting 
information about the anisotropy and boundaries of magnetic topologies
in solar flare regions. 

\section{CONCLUSIONS}

We analyzed the spatio-temporal evolution in the 155 largest solar 
flares (M and X-class) observed by GOES and AIA/SDO during the first
two years of the SDO mission. We fitted the radial expansion $r(t)$
of flare areas detected in the 335 \ang\ filter above some threshold
level with a generalized diffusion model that includes the classical
diffusion, anomalous diffusion, and the logistic growth limit.
The major results and conclusions are:

\begin{enumerate}
\item{The flare area in all events can be fitted with a radial expansion 
	model $r(t)$ that consists of an initial acceleration phase with 
	exponential growth and a deceleration phase that follows 
	anomalous diffusion, $r(t) \propto \kappa t^{\beta/2}$, with
	$\beta=0.53\pm0.27$ mostly falling into the sub-diffusive regime.
	The most extreme cases range from logistic growth ($\beta=0.04$)
	to super-diffusion ($\beta=1.35$). The sub-diffusive characteristics
	is likely to reflect the anisotropic propagation of energy release
	in a magnetically dominated plasma. The limit of logistic growth
	indicates the times when the boundaries of energy release regions
	are reached.}

\item{The diffusion coefficient $\kappa = 53\pm23$ km s$^{-\delta/2}$, which
	corresponds to an area diffusion constant of 
	$D \approx 300-1600$ km$^2$ s$^{-1}$ is found to be significantly
	faster than diffusion processes measured in the photosphere
	($D \approx 100-600$ km s$^{-1}$) and cannot be explained with
	crossfield diffusion in the corona, which is strongly inhibited 
	by the low plasma-$\beta$ parameter.}

\item{The average diffusion speed during flares is measured in the range
	of $v \approx 5-100$ km s$^{-1}$, with maximum speeds of
	$v_{max} \approx 10-500$ km s$^{-1}$, which is slower than the
	sound speed of $c_s \approx 500-900$ km s$^{-1}$ expected in
	flares with temperatures of $T_e \approx 10-30$ MK, but is
	compatible with the hard X-ray footpoint motion along the neutral 
	line during flares, and thus is likely to represent the mean
	propagation speed of subsequently triggered magnetic reconnection
	sites.}

\item{The fractal-diffusive self-organized criticality model (FD-SOC)
	describes the diffusive flare progression in a fractal geometry
	and predicts powerlaw distributions with slopes of
	$\alpha_L=3.0$ for length scales $L$,
	$\alpha_T=1.1-2.0$ for time scales $T$, and
	$\alpha_P=1.83$ for peak fluxes $P$,
	based on the observed diffusion index range of $\beta \approx
	0.1-1.0$, the observed flux-volume scaling of $F_{335} \propto
	(dV/dt)^{0.8}$, in a 3D-space geometry with fractal dimension 
	$D_3=2.0$.
	Our observations yield $\alpha_L=2.0$, $\alpha_T=2.2$, and
	$\alpha_P=1.34$, which is not inconsistent with the FD-SOC model,
	given the large uncertainties of the powerlaw slopes determined
 	in a small statistical sample. Nevertheless, the evolutionary
	fits $r(t) \propto t^{\beta/2}$ of individual flares confirm 
	the diffusive property of the FD-SOC model for the case of
	solar flares with a high degree of accuracy, which has been 
	previously predicted based on cellular automaton simulations.}
\end{enumerate}

Thus, the major accomplishment of this study is the experimental proof
of a diffusive flare expansion process that has been predicted by the
fractal-diffusive self-organized criticality (FD-SOC) model. However, 
while the classical FD-SOC model assumed classical diffusion ($\beta=1$),
the observational results from AIA/SDO revealed mostly sub-diffusion
$(\beta \lapprox 1)$, down to the limit of logistic growth 
($\beta \gapprox 0$). 
Future studies may extend the statistics of spatio-temporal parameters and
allow us to test the predicted size distributions of SOC models
with higher accuracy and establish physical scaling laws for flares
that are based on the fundamental geometric parameters of space and time. 

\acknowledgements
The author acknowledges helpful discussions and software support
of the AIA/SDO team. This work was partially supported by 
NASA contract NNX11A099G ``Self-organized criticality in solar physics''
and NASA contract NNG04EA00C of the SDO/AIA instrument to LMSAL.

\section*{Appendix A: Generalized Fractal-Diffusive Self-Organized 
Criticality Model}

We generalize the {\sl fractal-diffusive self-organized criticality model
(FD-SOC)} described in Aschwanden (2012) in two respects, (1) regarding
anomalous diffusion, and (2) for arbitrary (wavelength-dependent)
flux-volume scaling.
In the original FD-SOC model, classical diffusion was assumed, i.e., 
a scaling of $L \propto T^{1/2}$ between the length scale $L$ and 
time duration $T$, which we generalize for anomalous diffusion 
(with diffusion index $\beta$),
$$
	L \propto T^{\beta/2} \ .
	\eqno(A1)
$$
which includes also the case of classical diffusion ($\beta=1$).
The second generalization is the the scaling of an observed flux
$F_\lambda$ in a given wavelength $\lambda$ to the emissive volume 
rate of change $dV_S/dt$,
$$
	F_{\lambda} \approx \left( {dV_S \over dt} \right)^\gamma \ ,
	\eqno(A2)
$$
where $\gamma=1$ was assumed to be unity in the original FD-SOC model,
but can now be wavelength-dependent and accomodate some astrophysical
scaling law of the underlying emission and absorption process at
wavelength $\lambda$.

The FD-SOC model describes the inhomogeneity of SOC avalanches with a
fractal structure with Hausdorff dimension $D_S$ in Euclidean space 
with dimension $S$. The instantaneous avalanche volume 
$dV_S/dt$ is then characterized by  
$$
	{dV_S \over dt} \propto L^{D_S} \ ,
	\eqno(A3)
$$
and the mean fractal dimension $D_S$ can be estimated from the arithmetic
mean of the minimum and maximum possible value,
$$
	D_S \approx {(1+S) \over 2} \ ,
	\eqno(A4)
$$
yielding a mean value of $D_3=2.0$ for the 3D Euclidean space $(S=3)$.
Combining Eqs.~(A1-A3) yields then the correlation
of the instantaneous energy dissipation volume $(dV_S/dt)$ 
and the time duration $T$,
$$
	F_\lambda \propto \left({dV_S \over dt}\right)^\gamma 
	\propto L^{D_S \gamma} \propto T^{D_S \gamma \beta / 2} \ .
	\eqno(A5)
$$
This flux quantity $F_\lambda$ is generally strongly fluctuating, and the peak
dissipation rate $P_\lambda=max(F_\lambda)$ can be estimated from the maximum
fractal dimension $D_S \lapprox S$, 
$$
	P_\lambda \propto \left({dV_S \over dt}\right)^\gamma 
	\propto L^{S \gamma} \propto T^{S \gamma \beta / 2} \ .
	\eqno(A6)
$$
A more steady quantity is the time-integrated volume $V_S$ of a
SOC avalanche, which can be obtained from the time-integration of the
instantaneous volume rate of change $(dV_S/dt)$, and be associated with
the fluence $E_\lambda$ (or time-integrated flux $E_\lambda =
\int F_\lambda(t) dt$),  
$$
	E_\lambda \propto \int_0^T F_\lambda(t) \ dt 
	\propto \int_0^T t^{D_S \gamma \beta/2} \ dt 
	\propto T^{1+(D_S \gamma \beta / 2)} \ .
	\eqno(A7)
$$
After we obtained the scaling between the parameters $L, T, F_\lambda, 
P_\lambda$, and $E_\lambda$,
we can calculate the powerlaw slopes of their size distributions.
The basic reference is the distribution function $N(L)$ of spatial
scales, which be reciprocal to the size in a slowly-driven homogeneous 
SOC system that is scale-free and where avalanches can occur in any size. 
The relative probability is then (Aschwanden 2012),
$$
	N(L) \propto V_S^{-1} \propto L^{-S} \ .
	\eqno(A8)
$$
All other frequency distributions can then be derived by substituting
the corresponding parameter correlations given in Eq.~(A1-A7),
First we can
calculate the occurrence frequency distribution of avalanche time
scales, by using the diffusive boundary propagation relationship
$L(T) \propto T^{\beta/2}$ (Eq.~A1), by substituting the variable $T$
for $L$ in the distribution $N(L)$ (Eq.~A8),
$$
        N(T) dT = N(L[T]) \left| {dL \over dT} \right| dT
        \propto T^{-[1+(S-1) \beta /2]} \ dT \ .
	\eqno(A9)
$$
Subsequently we can derive the occurrence frequency distribution function
$N(F)$ for the instantaneous energy dissipation rate $F$
using the relationship $F_\lambda(T) \propto T^{D_S \gamma \beta/2}$ (Eq.~A5),
$$
        N(F_\lambda) dF_\lambda = N(T[F_\lambda]) \left| {dT \over dF_\lambda} 
	\right| dF_\lambda
        \propto F_\lambda^{-[1+(S-1)/D_S \gamma]} \ dF \ ,
	\eqno(A10)
$$
the occurrence frequency distribution function of the peak energy
dissipation rate $P$ using the relationship
relationship $P_\lambda(T) \propto T^{S \gamma \beta/2}$ (Eq.~A6),
$$
        N(P_\lambda) dP_\lambda = N(T[P_\lambda]) \left| 
	{dT \over dP_\lambda} \right| dP_\lambda
        \propto P_\lambda^{-[1+(S-1)/S \gamma]} \ dP \ ,
	\eqno(A11)
$$
and the occurrence frequency distribution function
$N(E_\lambda)$ for the total energy $E_\lambda$ using the relationship
$E_\lambda(T) \propto T^{(1+D_S \gamma \beta/2)}$ (Eq.~A7),
$$
        N(E_\lambda) dE_\lambda = N(T[E_\lambda]) \left| 
	{dT \over dE_\lambda} \right| dE_\lambda
        \propto E_\lambda^{-[1+(S-1)/(D_S \gamma +2/\beta)]} \ dE \ .
	\eqno(A12)
$$
This derivation yields naturally powerlaw functions for all
parameters $L$, $T$, $F_\lambda$, $P_\lambda$, and $E_\lambda$, 
which are the hallmarks of SOC systems.
In summary, if we denote the occurrence frequency distributions
$N(x)$ of a parameter $x$ with a powerlaw distribution with power index
$\alpha_x$,
$$
        N(x) dx \propto x^{-\alpha_x} \ dx \ ,
	\eqno(A13)
$$
we have the following powerlaw coefficients $\alpha_x$ for the parameters
$x=L, T, F_\lambda, P_\lambda$, and $E_\lambda$,
$$
        \begin{array}{ll}
        \alpha_L &=  S                    \\
        \alpha_T &=  1+(S-1) \beta/2      \\
        \alpha_F &=  1+(S-1)/(D_S \gamma) \\
        \alpha_P &=  1+(S-1)/(S \gamma)   \\
        \alpha_E &=  1+(S-1)/(D_S \gamma+2/\beta) \\
        \end{array} \ .
	\eqno(A14)
$$
Note, that the powerlaw slopes $\alpha_L$, $\alpha_F$ and $\alpha_P$ do not 
depend on the diffusion index $\beta$, and thus are identical for classical
or anomalous diffusion, while $\alpha_T$ and $\alpha_E$ depend on
the diffusion index $\beta$.

\clearpage


\begin{deluxetable}{rrrrrrrrrrrrr}
 \tabletypesize{\normalsize}
\tabletypesize{\footnotesize}
\tablecaption{Catalog of analyzed M and X-class flare events and best-fit 
model parameters: length scale $L$(Mm), diffusion coefficient $\kappa$
(km s$^{-\beta/2})$, diffusion index $\beta$, and goodness-of-fit $q_{fit}$.}
\tablewidth{0pt}
\tablehead{
\colhead{Nr}&
\colhead{Observation}&
\colhead{Start}&
\colhead{Peak}&
\colhead{End}&
\colhead{Duration}&
\colhead{GOES}&
\colhead{NOAA}&
\colhead{Heliogr.}&
\colhead{Length}&
\colhead{Diff.}&
\colhead{Diff.}&
\colhead{Fit}\\
\colhead{  }&
\colhead{date}&
\colhead{time}&
\colhead{time}&
\colhead{time}&
\colhead{T(s)}&
\colhead{class}&
\colhead{AR}&
\colhead{position}&
\colhead{L(Mm)}&
\colhead{coeff. $\kappa$}&
\colhead{index $\beta$}&
\colhead{$q_{fit}$}}
\startdata
   1 &2010-06-12 &00:30 &00:57 &01:02 & 1920 &M2.0 &11081 &N23W47 & 12 & 65$\pm$25 &0.60$\pm$0.27 &2.9\% \\
   2 &2010-06-13 &05:30 &05:39 &05:44 &  840 &M1.0 &11079 &S24W82 &  8 & 35$\pm$ 4 &0.81$\pm$0.11 &1.8\% \\
   3 &2010-08-07 &17:55 &18:24 &18:47 & 3120 &M1.0 &11093 &N14E37 & 40 &107$\pm$ 4 &0.36$\pm$0.02 &2.2\% \\
   4 &2010-10-16 &19:07 &19:12 &19:15 &  480 &M2.9 &11112 &S20W26 & 17 & 86$\pm$18 &0.55$\pm$0.03 &2.6\% \\
   5 &2010-11-04 &23:30 &23:58 &00:12 & 2520 &M1.6 &11121 &S20E85 & 12 & 34$\pm$ 0 &0.23$\pm$0.05 &1.9\% \\
   6 &2010-11-05 &12:43 &13:29 &14:06 & 4980 &M1.0 &11121 &S20E75 & 14 & 38$\pm$ 2 &0.58$\pm$0.06 &1.6\% \\
   7 &2010-11-06 &15:27 &15:36 &15:44 & 1020 &M5.4 &11121 &S20E58 & 14 & 68$\pm$ 6 &0.53$\pm$0.01 &1.7\% \\
   8 &2011-01-28 &00:44 &01:03 &01:10 & 1560 &M1.3 &11149 &N16W88 & 11 & 30$\pm$ 2 &0.12$\pm$0.06 &3.3\% \\
   9 &2011-02-09 &01:23 &01:31 &01:35 &  720 &M1.9 &11153 &N16W70 & 14 & 46$\pm$12 &0.23$\pm$0.09 &3.4\% \\
  10 &2011-02-13 &17:28 &17:38 &17:47 & 1140 &M6.6 &11158 &S20E05 & 21 & 74$\pm$12 &0.46$\pm$0.04 &2.6\% \\
  11 &2011-02-14 &17:20 &17:26 &17:32 &  720 &M2.2 &11158 &S20E04 & 23 & 66$\pm$11 &0.15$\pm$0.01 &2.2\% \\
  12 &2011-02-15 &01:44 &01:56 &02:06 & 1320 &X2.2 &11158 &S20W10 & 33 &114$\pm$ 3 &0.66$\pm$0.28 &2.9\% \\
  13 &2011-02-16 &01:32 &01:39 &01:46 &  840 &M1.0 &11158 &S20W24 & 12 & 46$\pm$ 5 &0.56$\pm$0.04 &2.5\% \\
  14 &2011-02-16 &07:35 &07:44 &07:55 & 1200 &M1.1 &11161 &S19W29 & 10 & 41$\pm$ 4 &0.66$\pm$0.01 &2.5\% \\
  15 &2011-02-16 &14:19 &14:25 &14:29 &  600 &M1.6 &11158 &S21W30 & 16 & 41$\pm$ 4 &0.10$\pm$0.01 &1.1\% \\
  16 &2011-02-18 &09:55 &10:11 &10:15 & 1200 &M6.6 &11158 &S21W55 & 13 & 81$\pm$13 &0.57$\pm$0.05 &4.2\% \\
  17 &2011-02-18 &10:23 &10:26 &10:37 &  840 &M1.0 &11162 &N22E10 & 14 & 41$\pm$ 4 &0.25$\pm$0.03 &2.6\% \\
  18 &2011-02-18 &12:59 &13:03 &13:06 &  420 &M1.4 &11158 &S20W70 &  7 & 47$\pm$ 5 &0.81$\pm$0.03 &1.8\% \\
  19 &2011-02-18 &14:00 &14:08 &14:15 &  900 &M1.0 &11162 &N22E10 & 14 & 40$\pm$ 5 &0.17$\pm$0.00 &2.0\% \\
  20 &2011-02-18 &20:56 &21:04 &21:14 & 1080 &M1.3 &11162 &N22E10 &  9 & 48$\pm$ 2 &0.81$\pm$0.09 &1.8\% \\
  21 &2011-02-24 &07:23 &07:35 &07:42 & 1140 &M3.5 &11163 &N14E87 & 17 & 77$\pm$17 &0.88$\pm$0.12 &2.5\% \\
  22 &2011-02-28 &12:38 &12:52 &13:03 & 1500 &M1.1 &11164 &N28E39 & 20 & 55$\pm$ 6 &0.28$\pm$0.02 &2.3\% \\
  23 &2011-03-07 &05:00 &05:13 &05:19 & 1140 &M1.2 &11164 &N23E47 & 21 & 71$\pm$11 &0.58$\pm$0.03 &2.2\% \\
  24 &2011-03-07 &07:49 &07:54 &07:56 &  420 &M1.5 &11165 &S18W75 &  9 & 38$\pm$15 &0.34$\pm$0.19 &2.7\% \\
  25 &2011-03-07 &07:59 &08:07 &08:15 &  960 &M1.4 &11164 &N27W46 & 23 & 63$\pm$ 6 &0.18$\pm$0.02 &2.0\% \\
  26 &2011-03-07 &09:14 &09:20 &09:28 &  840 &M1.8 &11164 &S17W77 &  5 & 31$\pm$15 &0.92$\pm$0.33 &3.0\% \\
  27 &2011-03-07 &13:45 &14:30 &14:56 & 4260 &M1.9 &11166 &N11E21 & 30 & 72$\pm$ 6 &0.42$\pm$0.04 &2.1\% \\
  28 &2011-03-07 &19:43 &20:12 &20:58 & 4500 &M3.7 &11164 &N30W48 & 51 &117$\pm$10 &0.47$\pm$0.08 &1.2\% \\
  29 &2011-03-07 &21:45 &21:50 &21:55 &  600 &M1.5 &11165 &S17W82 & 10 & 25$\pm$ 1 &0.10$\pm$0.01 &2.0\% \\
  30 &2011-03-08 &02:24 &02:29 &02:32 &  480 &M1.3 &11165 &S18W80 &  9 & 27$\pm$ 3 &0.26$\pm$0.03 &2.0\% \\
  31 &2011-03-08 &03:37 &03:58 &04:20 & 2580 &M1.5 &11171 &S21E72 & 32 & 90$\pm$ 5 &0.79$\pm$0.07 &1.7\% \\
  32 &2011-03-08 &10:35 &10:44 &10:55 & 1200 &M5.3 &11165 &S17W88 & 19 & 59$\pm$ 9 &0.38$\pm$0.03 &3.3\% \\
  33 &2011-03-08 &18:08 &18:28 &18:41 & 1980 &M4.4 &11165 &S17W88 & 13 & 37$\pm$ 7 &0.52$\pm$0.15 &2.7\% \\
  34 &2011-03-08 &19:46 &20:16 &21:19 & 5580 &M1.4 &11165 &N23W43 & 26 & 60$\pm$ 3 &0.50$\pm$0.09 &2.0\% \\
  35 &2011-03-09 &10:35 &11:07 &11:21 & 2760 &M1.7 &11166 &N08W11 & 21 & 56$\pm$ 6 &0.63$\pm$0.04 &3.2\% \\
  36 &2011-03-09 &13:17 &14:02 &14:13 & 3360 &M1.7 &11166 &S26W78 & 12 & 27$\pm$ 7 &0.73$\pm$0.12 &2.8\% \\
  37 &2011-03-09 &23:13 &23:23 &23:29 &  960 &X1.5 &11166 &N08W11 & 34 &123$\pm$17 &0.56$\pm$0.03 &1.9\% \\
  38 &2011-03-10 &22:34 &22:41 &22:49 &  900 &M1.1 &11166 &N08W25 &  8 & 42$\pm$10 &1.14$\pm$0.01 &2.3\% \\
  39 &2011-03-12 &04:33 &04:43 &04:48 &  900 &M1.3 &11166 &N07W35 & 16 & 53$\pm$ 9 &0.35$\pm$0.03 &1.7\% \\
  40 &2011-03-14 &19:30 &19:52 &19:54 & 1440 &M4.2 &11169 &N16W49 & 17 & 54$\pm$ 6 &0.26$\pm$0.04 &3.4\% \\
  41 &2011-03-15 &00:18 &00:22 &00:24 &  360 &M1.0 &11169 &N11W83 &  9 & 27$\pm$ 5 &0.17$\pm$0.06 &2.7\% \\
  42 &2011-03-23 &02:03 &02:17 &02:24 & 1260 &M1.4 &11176 &S25W84 & 18 & 58$\pm$ 8 &0.45$\pm$0.02 &1.4\% \\
  43 &2011-03-24 &12:01 &12:07 &12:11 &  600 &M1.0 &11176 &S15E43 & 15 & 48$\pm$ 5 &0.26$\pm$0.03 &2.3\% \\
  44 &2011-03-25 &23:08 &23:22 &23:30 & 1320 &M1.0 &11176 &S12E26 & 20 & 59$\pm$ 2 &0.39$\pm$0.06 &1.9\% \\
  45 &2011-04-15 &17:02 &17:12 &17:28 & 1560 &M1.3 &11190 &N13W24 & 13 & 43$\pm$ 6 &0.42$\pm$0.04 &1.4\% \\
  46 &2011-04-22 &04:35 &04:57 &05:14 & 2340 &M1.8 &11195 &S17E42 & 20 & 60$\pm$12 &0.77$\pm$0.14 &2.0\% \\
  47 &2011-04-22 &15:47 &15:53 &16:11 & 1440 &M1.2 &11195 &S18E36 & 20 & 57$\pm$ 5 &0.36$\pm$0.03 &1.5\% \\
  48 &2011-05-28 &21:09 &21:50 &22:01 & 3120 &M1.1 &11226 &S21E70 & 14 & 37$\pm$ 5 &0.72$\pm$0.08 &1.9\% \\
  49 &2011-05-29 &10:08 &10:33 &11:08 & 3600 &M1.4 &11226 &S20E64 & 24 & 68$\pm$ 6 &0.93$\pm$0.14 &2.5\% \\
  50 &2011-06-07 &06:16 &06:41 &06:59 & 2580 &M2.5 &11226 &S22W53 & 33 & 85$\pm$ 7 &0.30$\pm$0.02 &1.5\% \\
  51 &2011-06-14 &21:36 &21:47 &22:10 & 2040 &M1.3 &11236 &N14E77 & 20 & 55$\pm$ 4 &0.37$\pm$0.02 &1.7\% \\
  52 &2011-07-27 &15:48 &16:07 &16:22 & 2040 &M1.1 &11260 &N19E38 & 19 & 63$\pm$ 8 &0.55$\pm$0.15 &2.7\% \\
  53 &2011-07-30 &02:04 &02:09 &02:12 &  480 &M9.3 &11261 &N14E35 & 26 &139$\pm$20 &0.62$\pm$0.01 &2.7\% \\
  54 &2011-08-02 &05:19 &06:19 &06:48 & 5340 &M1.4 &11261 &N17W12 & 28 & 74$\pm$ 7 &0.68$\pm$0.05 &1.7\% \\
  55 &2011-08-03 &03:08 &03:37 &03:51 & 2580 &M1.1 &11261 &N15W28 & 13 & 37$\pm$ 4 &0.24$\pm$0.06 &2.6\% \\
  56 &2011-08-03 &04:29 &04:32 &04:35 &  360 &M1.7 &11263 &N16E10 & 21 & 73$\pm$ 7 &0.29$\pm$0.01 &1.0\% \\
  57 &2011-08-03 &13:17 &13:48 &14:10 & 3180 &M6.0 &11261 &N17W30 & 28 & 78$\pm$ 3 &0.72$\pm$0.04 &3.0\% \\
  58 &2011-08-04 &03:41 &03:57 &04:04 & 1380 &M9.3 &11261 &N16W38 & 32 & 92$\pm$16 &0.28$\pm$0.02 &1.9\% \\
  59 &2011-08-08 &18:00 &18:10 &18:18 & 1080 &M3.5 &11263 &N15W62 & 17 & 57$\pm$ 6 &0.47$\pm$0.01 &1.5\% \\
  60 &2011-08-09 &03:19 &03:54 &04:08 & 2940 &M2.5 &11263 &N17W69 & 12 & 39$\pm$ 5 &1.35$\pm$0.03 &2.1\% \\
  61 &2011-08-09 &07:48 &08:05 &08:08 & 1200 &X6.9 &11263 &N14W69 & 29 & 95$\pm$12 &0.19$\pm$0.00 &3.2\% \\
  62 &2011-09-04 &11:21 &11:45 &11:50 & 1740 &M3.2 &11286 &N18W84 & 10 & 38$\pm$ 1 &0.50$\pm$0.14 &3.3\% \\
  63 &2011-09-05 &04:08 &04:28 &04:32 & 1440 &M1.6 &11286 &N18W87 & 13 & 46$\pm$ 0 &0.55$\pm$0.12 &2.0\% \\
  64 &2011-09-05 &07:27 &07:58 &08:06 & 2340 &M1.2 &11286 &N18W87 &  7 & 22$\pm$ 2 &1.33$\pm$0.29 &2.3\% \\
  65 &2011-09-06 &01:35 &01:50 &02:05 & 1800 &M5.3 &11283 &N13W07 & 21 & 60$\pm$ 8 &0.37$\pm$0.05 &0.9\% \\
  66 &2011-09-06 &22:12 &22:20 &22:24 &  720 &X2.1 &11283 &N14W18 & 29 & 87$\pm$ 6 &0.18$\pm$0.03 &3.4\% \\
  67 &2011-09-07 &22:32 &22:38 &22:44 &  720 &X1.8 &11283 &N14W31 & 36 &101$\pm$15 &0.17$\pm$0.02 &1.2\% \\
  68 &2011-09-08 &15:32 &15:46 &15:52 & 1200 &M6.7 &11283 &N14W41 & 27 & 80$\pm$14 &0.32$\pm$0.08 &1.8\% \\
  69 &2011-09-09 &06:01 &06:11 &06:17 &  960 &M2.7 &11283 &N14W48 & 16 & 73$\pm$21 &0.85$\pm$0.30 &1.8\% \\
  70 &2011-09-09 &12:39 &12:49 &12:56 & 1020 &M1.2 &11283 &N15W50 & 15 & 45$\pm$ 5 &0.35$\pm$0.01 &1.1\% \\
  71 &2011-09-10 &07:18 &07:40 &07:56 & 2280 &M1.1 &11283 &N14W64 & 15 & 47$\pm$ 6 &0.91$\pm$0.02 &1.0\% \\
  72 &2011-09-21 &12:04 &12:23 &12:45 & 2460 &M1.8 &11301 &N15E88 &  6 & 18$\pm$ 3 &0.48$\pm$0.03 &2.6\% \\
  73 &2011-09-22 &09:53 &10:00 &10:09 &  960 &M1.1 &11302 &N24W55 & 12 & 52$\pm$ 9 &0.87$\pm$0.04 &3.1\% \\
  74 &2011-09-22 &10:29 &11:01 &11:44 & 4500 &X1.4 &11302 &N08E89 & 41 & 85$\pm$ 5 &0.99$\pm$0.13 &1.3\% \\
  75 &2011-09-23 &01:47 &01:59 &02:10 & 1380 &M1.6 &11295 &N24W64 & 14 & 59$\pm$10 &1.20$\pm$0.02 &2.3\% \\
  76 &2011-09-23 &21:54 &22:15 &22:34 & 2400 &M1.6 &11295 &N12E56 & 17 & 52$\pm$ 5 &0.72$\pm$0.08 &2.1\% \\
  77 &2011-09-23 &23:48 &23:56 &00:04 &  960 &M1.9 &11302 &N12E56 & 23 & 69$\pm$12 &0.32$\pm$0.03 &2.0\% \\
  78 &2011-09-24 &09:21 &09:40 &09:48 & 1620 &X1.9 &11302 &N13E61 & 23 & 60$\pm$ 9 &0.16$\pm$0.02 &2.6\% \\
  79 &2011-09-24 &12:33 &13:20 &14:10 & 5820 &M7.1 &11302 &N15E59 & 39 & 80$\pm$ 9 &1.05$\pm$0.11 &2.0\% \\
  80 &2011-09-24 &16:36 &16:59 &17:15 & 2340 &M1.7 &11295 &N23W87 & 12 & 42$\pm$ 4 &1.07$\pm$0.07 &1.2\% \\
  81 &2011-09-24 &17:19 &17:25 &17:31 &  720 &M3.1 &11302 &N13E54 &  8 & 35$\pm$ 9 &0.79$\pm$0.17 &3.1\% \\
  82 &2011-09-24 &17:59 &18:15 &18:24 & 1500 &M2.8 &11302 &N13E56 & 18 & 46$\pm$ 7 &0.16$\pm$0.01 &1.4\% \\
  83 &2011-09-24 &19:09 &19:21 &19:41 & 1920 &M3.0 &11302 &N12E42 & 26 & 78$\pm$ 8 &0.74$\pm$0.02 &1.8\% \\
  84 &2011-09-24 &20:29 &20:36 &20:42 &  780 &M5.8 &11302 &N13E52 & 11 & 47$\pm$ 9 &0.55$\pm$0.03 &3.8\% \\
  85 &2011-09-24 &21:23 &21:27 &21:32 &  540 &M1.2 &11303 &N13E52 &  5 & 22$\pm$15 &0.45$\pm$0.32 &4.1\% \\
  86 &2011-09-24 &23:45 &23:58 &00:09 & 1440 &M1.0 &11303 &S28W66 &  8 & 23$\pm$ 2 &0.43$\pm$0.16 &2.2\% \\
  87 &2011-09-25 &02:27 &02:33 &02:37 &  600 &M4.4 &11302 &N22W87 &  8 & 41$\pm$12 &0.99$\pm$0.00 &2.4\% \\
  88 &2011-09-25 &04:31 &04:50 &05:05 & 2040 &M7.4 &11302 &N13E50 & 30 & 90$\pm$ 4 &0.58$\pm$0.08 &2.4\% \\
  89 &2011-09-25 &08:46 &08:49 &08:52 &  360 &M3.1 &11302 &N13E45 & 11 & 45$\pm$ 6 &0.63$\pm$0.02 &1.7\% \\
  90 &2011-09-25 &09:25 &09:35 &09:53 & 1680 &M1.5 &11303 &S28W71 & 10 & 29$\pm$ 0 &0.86$\pm$0.20 &4.1\% \\
  91 &2011-09-25 &15:26 &15:33 &15:38 &  720 &M3.7 &11302 &N13E44 & 25 & 72$\pm$ 9 &0.19$\pm$0.02 &1.6\% \\
  92 &2011-09-25 &16:51 &16:58 &17:09 & 1080 &M2.2 &11303 &N12E41 & 11 & 39$\pm$ 2 &0.51$\pm$0.06 &2.7\% \\
  93 &2011-09-26 &05:06 &05:08 &05:13 &  420 &M4.0 &11302 &N12E34 & 17 & 64$\pm$13 &0.43$\pm$0.02 &2.9\% \\
  94 &2011-09-26 &14:37 &14:46 &15:02 & 1500 &M2.6 &11302 &N12E30 & 28 & 76$\pm$ 5 &0.28$\pm$0.00 &1.1\% \\
  95 &2011-09-28 &13:24 &13:28 &13:30 &  360 &M1.2 &11302 &N11E00 & 16 & 51$\pm$ 5 &0.25$\pm$0.02 &2.6\% \\
  96 &2011-09-30 &18:55 &19:06 &19:15 & 1200 &M1.0 &11305 &N09E03 & 18 & 64$\pm$ 7 &0.67$\pm$0.11 &2.6\% \\
  97 &2011-10-01 &08:56 &09:59 &10:17 & 4860 &M1.2 &11305 &N09W04 & 24 & 68$\pm$ 6 &0.84$\pm$0.44 &2.2\% \\
  98 &2011-10-02 &00:37 &00:50 &00:59 & 1320 &M3.9 &11305 &N10W13 & 18 & 59$\pm$ 8 &0.45$\pm$0.05 &1.6\% \\
  99 &2011-10-02 &17:19 &17:23 &17:26 &  420 &M1.3 &11302 &N10W55 & 12 & 41$\pm$ 3 &0.27$\pm$0.00 &2.2\% \\
 100 &2011-10-20 &03:10 &03:25 &03:44 & 2040 &M1.6 &11318 &N18W88 & 11 & 28$\pm$ 6 &0.27$\pm$0.14 &3.4\% \\
 101 &2011-10-21 &12:53 &13:00 &13:08 &  900 &M1.3 &11319 &N05W79 & 12 & 49$\pm$ 0 &0.60$\pm$0.13 &2.5\% \\
 102 &2011-10-22 &10:00 &11:10 &13:09 &11340 &M1.3 &11314 &N27W87 & 50 &112$\pm$23 &0.23$\pm$0.02 &1.7\% \\
 103 &2011-10-31 &14:55 &15:08 &15:27 & 1920 &M1.1 &11313 &N20E88 &  8 & 28$\pm$10 &0.45$\pm$0.29 &3.0\% \\
 104 &2011-10-31 &17:21 &18:08 &18:50 & 5340 &M1.4 &11313 &N21E88 &  8 & 20$\pm$ 6 &0.49$\pm$0.29 &1.3\% \\
 105 &2011-11-02 &21:52 &22:01 &22:19 & 1620 &M4.3 &11339 &N20E77 & 16 & 49$\pm$ 9 &0.43$\pm$0.14 &3.3\% \\
 106 &2011-11-03 &10:58 &11:11 &11:20 & 1320 &M2.5 &11339 &N20E70 & 12 & 34$\pm$ 1 &0.30$\pm$0.02 &1.3\% \\
 107 &2011-11-03 &20:16 &20:27 &20:32 &  960 &X1.9 &11339 &N21E64 & 20 & 63$\pm$ 9 &0.34$\pm$0.05 &2.2\% \\
 108 &2011-11-03 &23:28 &23:36 &23:44 &  960 &M2.1 &11339 &N20E62 & 12 & 45$\pm$ 2 &0.67$\pm$0.10 &2.1\% \\
 109 &2011-11-04 &20:31 &20:40 &20:46 &  900 &M1.0 &11339 &N19E47 & 12 & 57$\pm$16 &0.98$\pm$0.42 &2.1\% \\
 110 &2011-11-05 &03:08 &03:35 &03:58 & 3000 &M3.7 &11339 &N20E47 & 29 & 86$\pm$ 8 &0.48$\pm$0.13 &1.7\% \\
 111 &2011-11-05 &11:10 &11:21 &11:42 & 1920 &M1.1 &11339 &N19E41 & 15 & 47$\pm$10 &0.87$\pm$0.16 &2.1\% \\
 112 &2011-11-05 &20:31 &20:38 &20:54 & 1380 &M1.8 &11339 &N21E37 & 15 & 42$\pm$ 3 &0.34$\pm$0.01 &2.4\% \\
 113 &2011-11-06 &00:46 &01:03 &01:24 & 2280 &M1.2 &11339 &N20E34 & 14 & 38$\pm$ 4 &0.34$\pm$0.07 &1.7\% \\
 114 &2011-11-06 &06:14 &06:35 &06:41 & 1620 &M1.4 &11339 &N21E33 & 19 & 52$\pm$10 &0.19$\pm$0.03 &3.7\% \\
 115 &2011-11-09 &13:04 &13:35 &14:12 & 4080 &M1.1 &11342 &N24E35 & 39 & 91$\pm$19 &0.55$\pm$0.04 &2.2\% \\
 116 &2011-11-15 &09:03 &09:12 &09:23 & 1200 &M1.2 &11348 &N21W72 & 10 & 35$\pm$10 &0.52$\pm$0.18 &2.7\% \\
 117 &2011-11-15 &12:30 &12:43 &12:50 & 1200 &M1.9 &11346 &S19E32 & 15 & 41$\pm$ 7 &0.21$\pm$0.04 &2.6\% \\
 118 &2011-11-15 &22:27 &22:35 &22:42 &  900 &M1.1 &11348 &N18W81 &  9 & 22$\pm$ 2 &0.04$\pm$0.01 &1.4\% \\
 119 &2011-12-25 &18:11 &18:16 &18:20 &  540 &M4.0 &11387 &S22W26 & 17 & 60$\pm$14 &0.42$\pm$0.06 &3.2\% \\
 120 &2011-12-26 &02:13 &02:27 &02:36 & 1380 &M1.5 &11387 &S21W33 & 13 & 56$\pm$ 6 &1.00$\pm$0.06 &2.2\% \\
 121 &2011-12-26 &20:12 &20:30 &20:36 & 1440 &M2.3 &11387 &S21W44 & 11 & 41$\pm$11 &0.65$\pm$0.14 &2.1\% \\
 122 &2011-12-29 &13:40 &13:50 &14:01 & 1260 &M1.9 &11389 &S25E70 & 11 & 33$\pm$ 4 &0.43$\pm$0.05 &3.0\% \\
 123 &2011-12-29 &21:43 &21:51 &21:59 &  960 &M2.0 &11389 &S25E67 & 14 & 59$\pm$ 0 &0.98$\pm$0.14 &1.6\% \\
 124 &2011-12-30 &03:03 &03:09 &03:13 &  600 &M1.2 &11389 &S25E67 &  9 & 27$\pm$ 2 &0.30$\pm$0.13 &4.0\% \\
 125 &2011-12-31 &13:09 &13:15 &13:19 &  600 &M2.4 &11389 &S25E46 & 13 & 38$\pm$ 6 &0.25$\pm$0.06 &3.6\% \\
 126 &2011-12-31 &16:16 &16:26 &16:34 & 1080 &M1.5 &11389 &S25E42 & 15 & 51$\pm$ 3 &0.58$\pm$0.08 &2.9\% \\
 127 &2012-01-14 &13:14 &13:18 &13:20 &  360 &M1.4 &11401 &N14E88 &  8 & 21$\pm$ 5 &0.10$\pm$0.08 &3.0\% \\
 128 &2012-01-17 &04:41 &04:53 &05:07 & 1560 &M1.0 &11401 &N18E53 & 13 & 38$\pm$ 2 &0.52$\pm$0.02 &1.5\% \\
 129 &2012-01-18 &19:04 &19:12 &19:27 & 1380 &M1.7 &11401 &N17E32 & 16 & 59$\pm$11 &0.17$\pm$0.04 &2.5\% \\
 130 &2012-01-19 &13:44 &16:05 &17:50 &14760 &M3.2 &11402 &N32E27 & 38 & 76$\pm$11 &0.45$\pm$0.10 &1.9\% \\
 131 &2012-01-23 &03:38 &03:59 &04:34 & 3360 &M8.7 &11402 &N33W21 & 37 &111$\pm$ 9 &0.98$\pm$0.09 &1.2\% \\
 132 &2012-01-27 &17:37 &18:37 &18:56 & 4740 &X1.7 &11402 &N33W85 & 40 &114$\pm$ 8 &0.95$\pm$0.07 &1.7\% \\
 133 &2012-02-06 &19:31 &20:00 &20:17 & 2760 &M1.0 &11410 &N19W62 & 28 & 89$\pm$ 5 &0.54$\pm$0.14 &1.6\% \\
 134 &2012-03-02 &17:29 &17:46 &18:07 & 2280 &M3.3 &11429 &N19W62 & 10 & 36$\pm$ 1 &0.75$\pm$0.03 &1.6\% \\
 135 &2012-03-04 &10:29 &10:52 &12:16 & 6420 &M2.0 &11429 &N16E65 & 23 & 48$\pm$ 2 &0.48$\pm$0.03 &1.7\% \\
 136 &2012-03-05 &02:30 &04:09 &04:43 & 7980 &X1.1 &11429 &N19E58 & 32 & 74$\pm$ 5 &0.44$\pm$0.03 &1.7\% \\
 137 &2012-03-05 &19:10 &19:16 &19:21 &  660 &M2.1 &11429 &N16E45 & 10 & 32$\pm$ 2 &0.36$\pm$0.03 &2.3\% \\
 138 &2012-03-05 &19:27 &19:30 &19:32 &  300 &M1.8 &11429 &N16E45 &  7 & 26$\pm$ 5 &0.34$\pm$0.18 &3.0\% \\
 139 &2012-03-05 &22:26 &22:34 &22:42 &  960 &M1.3 &11429 &N16E43 &  9 & 33$\pm$ 3 &0.46$\pm$0.01 &2.3\% \\
 140 &2012-03-06 &00:22 &00:28 &00:31 &  540 &M1.3 &11429 &N16E42 &  8 & 27$\pm$ 5 &0.38$\pm$0.20 &3.1\% \\
 141 &2012-03-06 &01:36 &01:44 &01:50 &  840 &M1.2 &11429 &N16E41 &  8 & 28$\pm$ 3 &0.40$\pm$0.04 &2.3\% \\
 142 &2012-03-06 &04:01 &04:05 &04:08 &  420 &M1.0 &11429 &N16E39 &  9 & 47$\pm$ 6 &0.96$\pm$0.09 &2.0\% \\
 143 &2012-03-06 &07:52 &07:55 &08:00 &  480 &M1.0 &11429 &N17E40 &  8 & 41$\pm$ 2 &0.87$\pm$0.05 &2.5\% \\
 144 &2012-03-06 &12:23 &12:41 &12:54 & 1860 &M2.1 &11429 &N21E40 & 15 & 55$\pm$10 &1.07$\pm$0.16 &1.3\% \\
 145 &2012-03-06 &21:04 &21:11 &21:14 &  600 &M1.3 &11429 &N16E30 & 10 & 38$\pm$ 9 &0.59$\pm$0.11 &2.0\% \\
 146 &2012-03-06 &22:49 &22:53 &23:11 & 1320 &M1.0 &11429 &N19E32 & 11 & 35$\pm$ 8 &0.62$\pm$0.07 &1.8\% \\
 147 &2012-03-07 &00:02 &00:24 &00:40 & 2280 &X5.4 &11429 &N18E31 & 37 &100$\pm$12 &0.60$\pm$0.01 &1.9\% \\
 148 &2012-03-07 &01:05 &01:14 &01:23 & 1080 &X1.3 &11430 &N15E26 & 35 &105$\pm$ 6 &0.63$\pm$0.03 &0.8\% \\
 149 &2012-03-09 &03:22 &03:53 &04:18 & 3360 &M6.3 &11429 &N15W03 & 33 & 82$\pm$ 5 &0.50$\pm$0.01 &3.6\% \\
 150 &2012-03-10 &17:15 &17:44 &18:30 & 4500 &M8.4 &11429 &N17W24 & 30 & 77$\pm$ 5 &0.40$\pm$0.01 &1.2\% \\
 151 &2012-03-13 &17:12 &17:41 &18:25 & 4380 &M7.9 &11429 &N17W66 & 36 & 83$\pm$ 6 &0.73$\pm$0.01 &1.3\% \\
 152 &2012-03-14 &15:08 &15:21 &15:36 & 1680 &M2.8 &11432 &N13E05 & 18 & 63$\pm$ 6 &0.55$\pm$0.03 &2.0\% \\
 153 &2012-03-15 &07:23 &07:52 &08:08 & 2700 &M1.8 &11432 &N14W03 & 21 & 66$\pm$ 6 &0.61$\pm$0.03 &1.3\% \\
 154 &2012-03-17 &20:32 &20:39 &20:42 &  600 &M1.3 &11434 &S20W25 & 10 & 42$\pm$ 4 &0.51$\pm$0.02 &1.1\% \\
 155 &2012-03-23 &19:34 &19:40 &19:44 &  600 &M1.0 &11445 &S20W25 &  7 & 23$\pm$ 2 &0.25$\pm$0.01 &2.5\% \\
\enddata
\end{deluxetable}


\begin{deluxetable}{lrrrr}
\tabletypesize{\normalsize}
\tablecaption{Statistical results of observables and model parameters 
of 155 analyzed flare events during 2010/05/13-2011/03/31.}
\tablewidth{0pt}
\tablehead{
\colhead{Parameter}&
\colhead{Range}&
\colhead{Mean}&
\colhead{Median}&
\colhead{Powerlaw slope}}
\startdata
Flare duration (GOES rise time) $T(s)=$	
& 120-8460  				& $1030\pm1050$ & 660 		& 2.17		\\
AIA 335 time interval $T_{AIA}(s)=$	
& 336-10992 				& $1806\pm1579$ & 1296		&		\\
Length (flare radius) $L(Mm)=$	
& $5.5-51$				& $19\pm10$     & 16		& 1.96		\\
Peak flux GOES $F_{GOES}($W m$^{-2}$)=
& $(1-69) \times 10^{-5}$		& $(4.1\pm7.8) \times 10^{-5}$  & $7.8 \times 10^{-5}$ & 1.92 \\
Total flux AIA 335 \ang\ $F_{AIA}(DN/s)$=
& $(7.2-2462) \times 10^4$ 		& $(4.4\pm4.7) \times 10^6$     & $2.5\times 10^6$ & 1.34 \\
Mean velocity $v (km/s)=$	
& 3-103 				& $15\pm12$     & 11 		& 2.72		\\	
Maximum velocity $v_{max} (km/s)=$	
& 8-550 				& $80\pm85$     & 57 		& 1.83		\\	
Diffusion coefficient $\kappa ($km s$^{-\beta/2}$)= 
& 19-139 				& $56\pm24$	& 52		& 		\\ 
Diffusion index $\beta=$
& 0.04-1.35				& $0.53\pm0.27$ & 0.49		&		\\
Goodness-of-fit $q_{fit}=\Delta r/r_{max}$(\%)=
& 0.8-4.2 				& $2.2 \pm 0.7$ & 2.1		&		\\
\enddata
\end{deluxetable}

\clearpage

\begin{figure}
\plotone{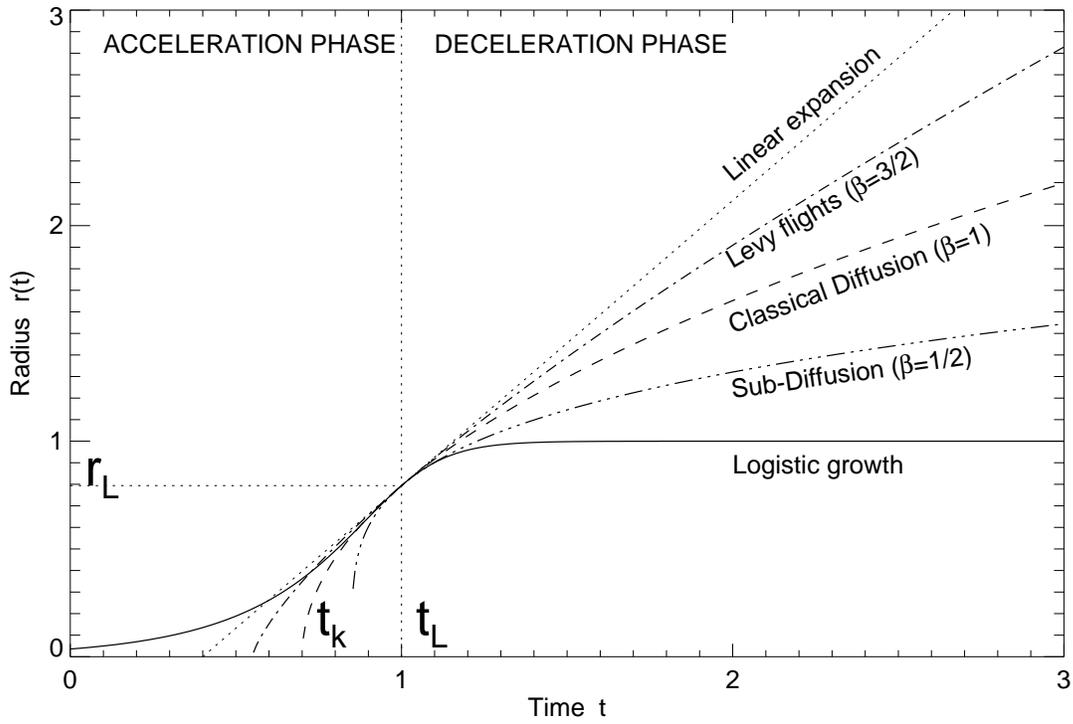}
\caption{Comparison of spatio-temporal evolution models:
Logistic growth with parameters $t_L=1.0, r_\infty=1.0, \tau_G=0.1$,
sub-diffusion ($\beta=1/2$), classical diffusion ($\beta=1$), 
L\'evy flights or super-diffusion ($\beta=3/2$), and linear expansion 
($r \propto t$).  All three curves intersect at $t=t_L$
and have the same speed $v=(dr/dt)$ at the intersection point at
time $t=t_L$.}
\end{figure}

\begin{figure}
\plotone{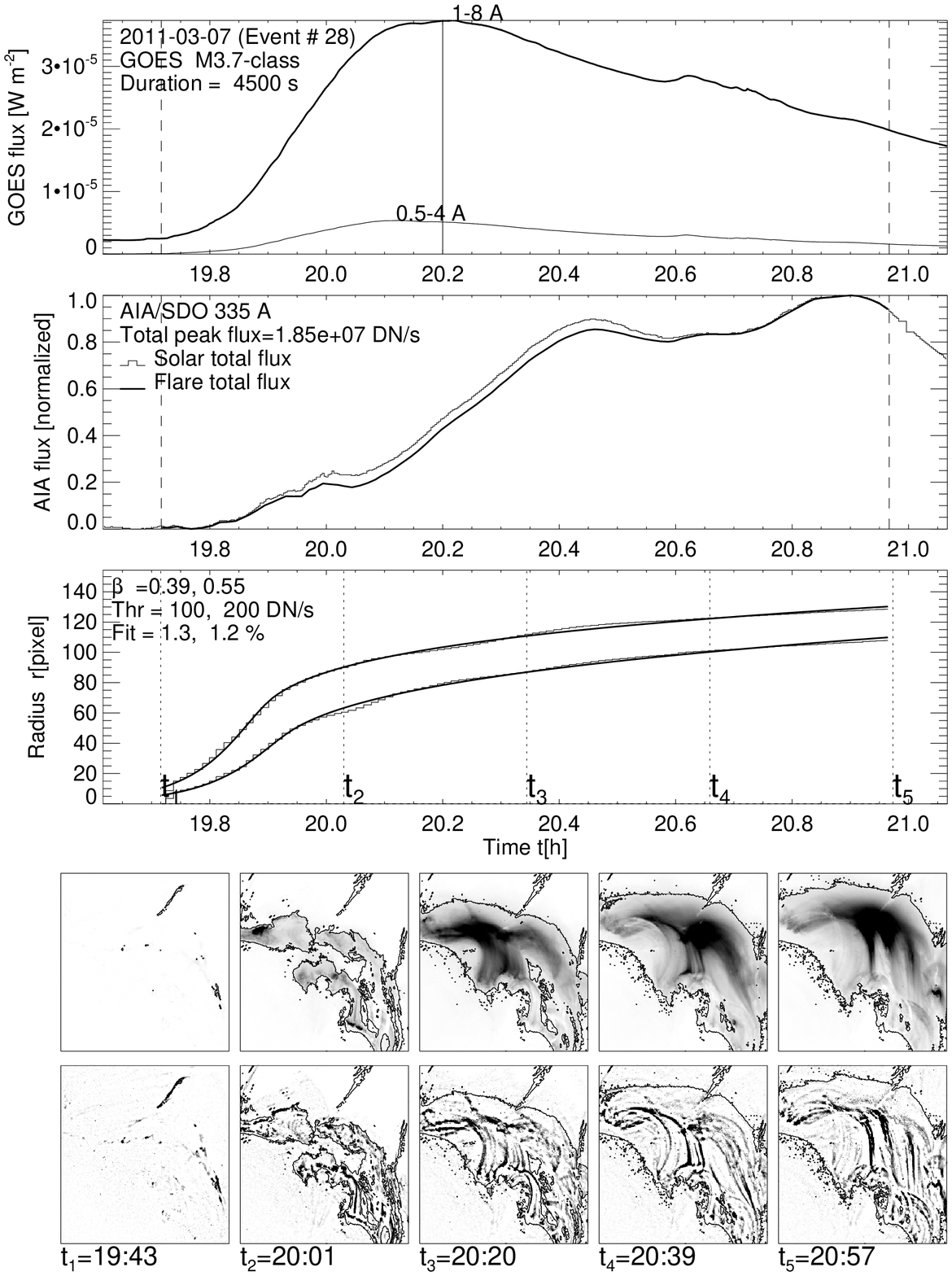}
\caption{The flare event $\#28$ was observed on 2011 Mar 7, 19:43-20:58 UT,
with AIA/SDO 335 \ang , with GOES time profiles (top panel), the
total flux (second panel), the spatio-temporal evolution of the radius 
$r(t)=\sqrt{A(t)/\pi}$ of the time-integrated flare area $A(t)$ 
for two thresholds, $F_{thresh}=100, 200$ DN/s (third panel; histogrammed), 
fitted with the anomalous diffusion model (third panel; solid curve),
and 5 snapshots of the baseline-subtracted flux (fourth row) and 
highpass-filtered flux (bottom row), with the threshold flux 
$F_{thresh}=100$ DN/s shown as contour.}
\end{figure}

\begin{figure}
\plotone{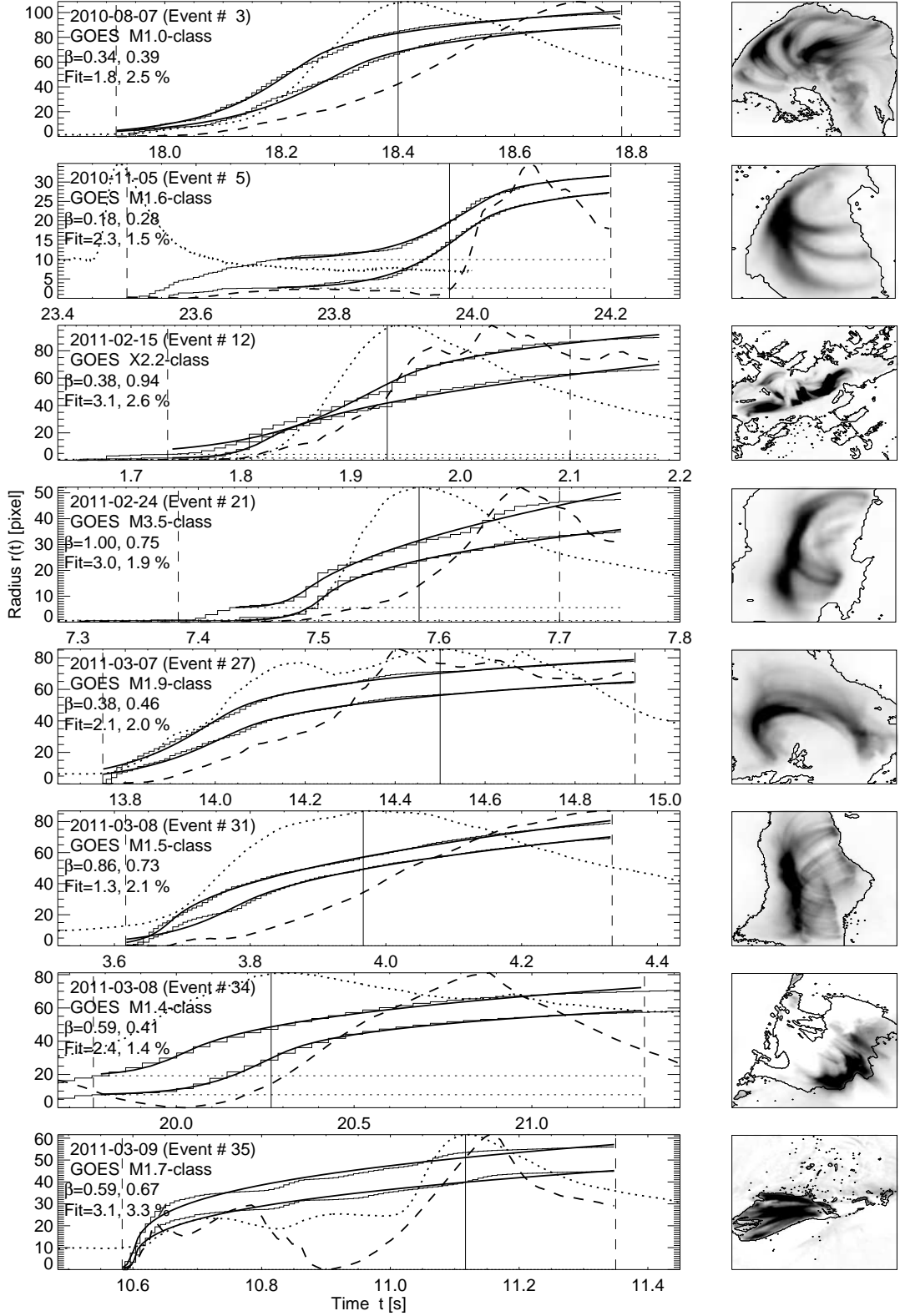}
\caption{A selection of analyzed flare events, showing the GOES 1-8 \ang\ 
time profile (dotted curve), the AIA 335 \ang\ total flux (dashed curve),
the measured radius of the flare area for two flux thresholds (histograms),
the best-fit diffusion model (solid curve), and preflare-subtracted AIA
335 image at the end of the flare, with the contours of the flux threshold 
of $F_{thresh}=100$ DN/s indicated (right panels).}
\end{figure}

\plotone{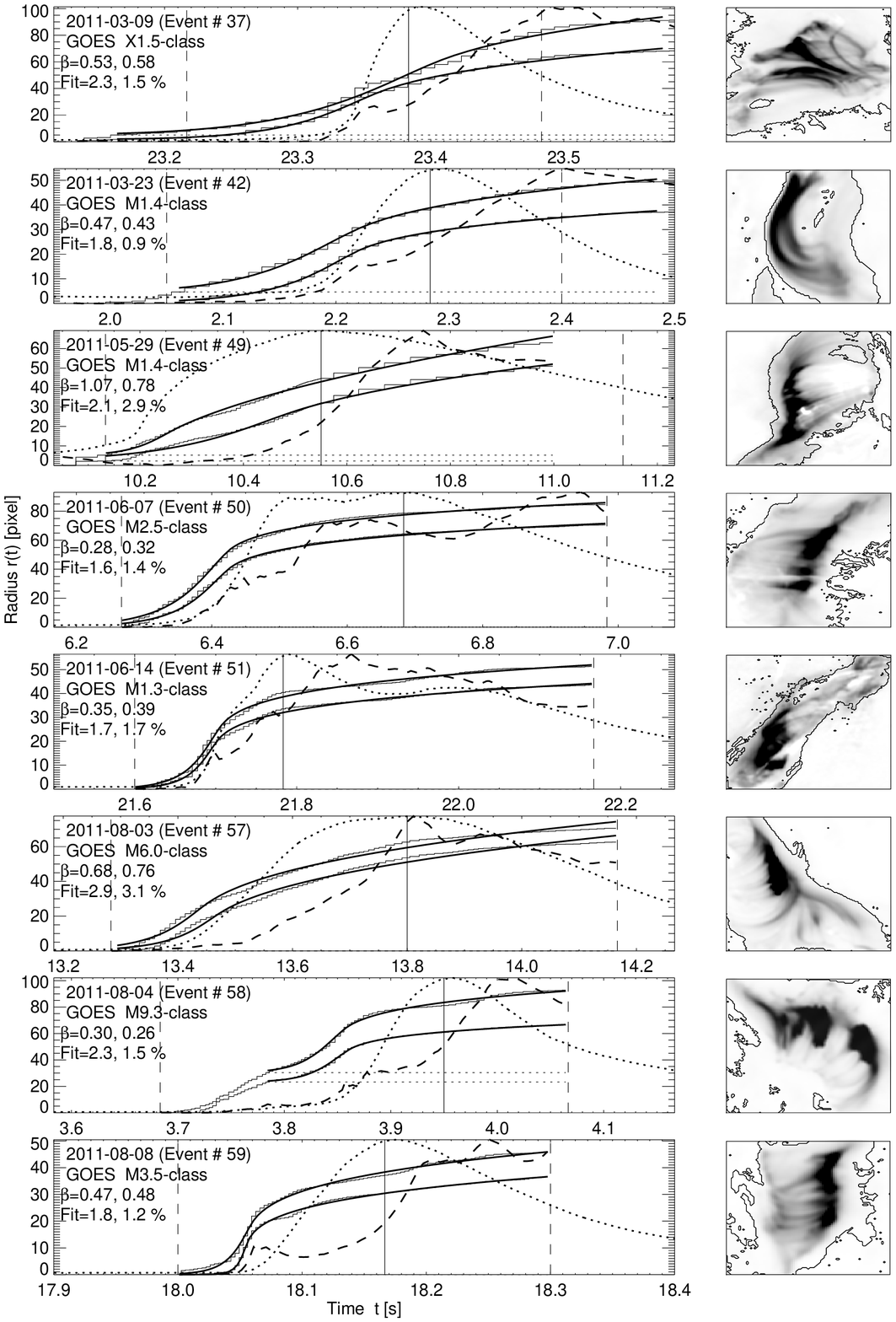}
\plotone{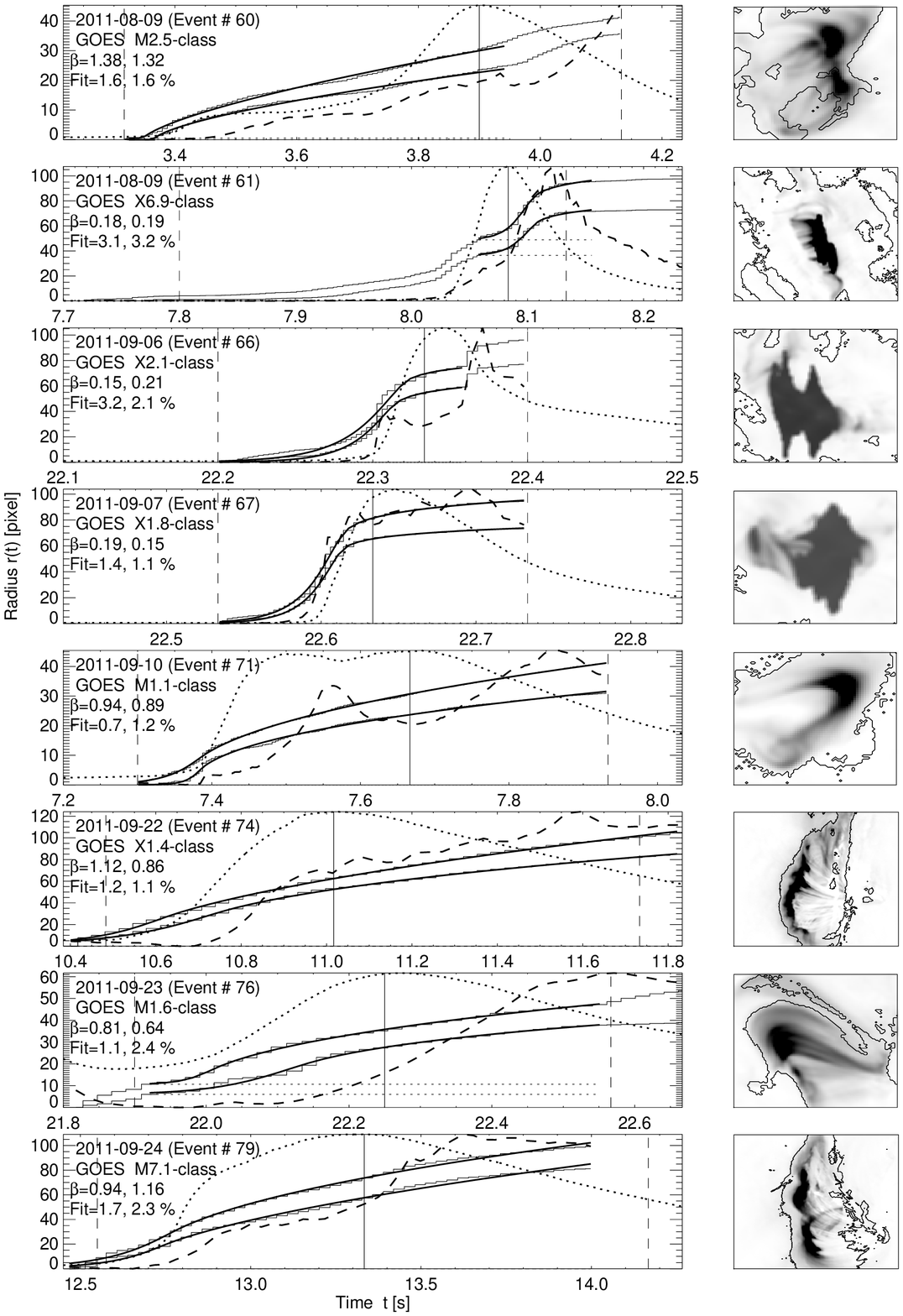}
\plotone{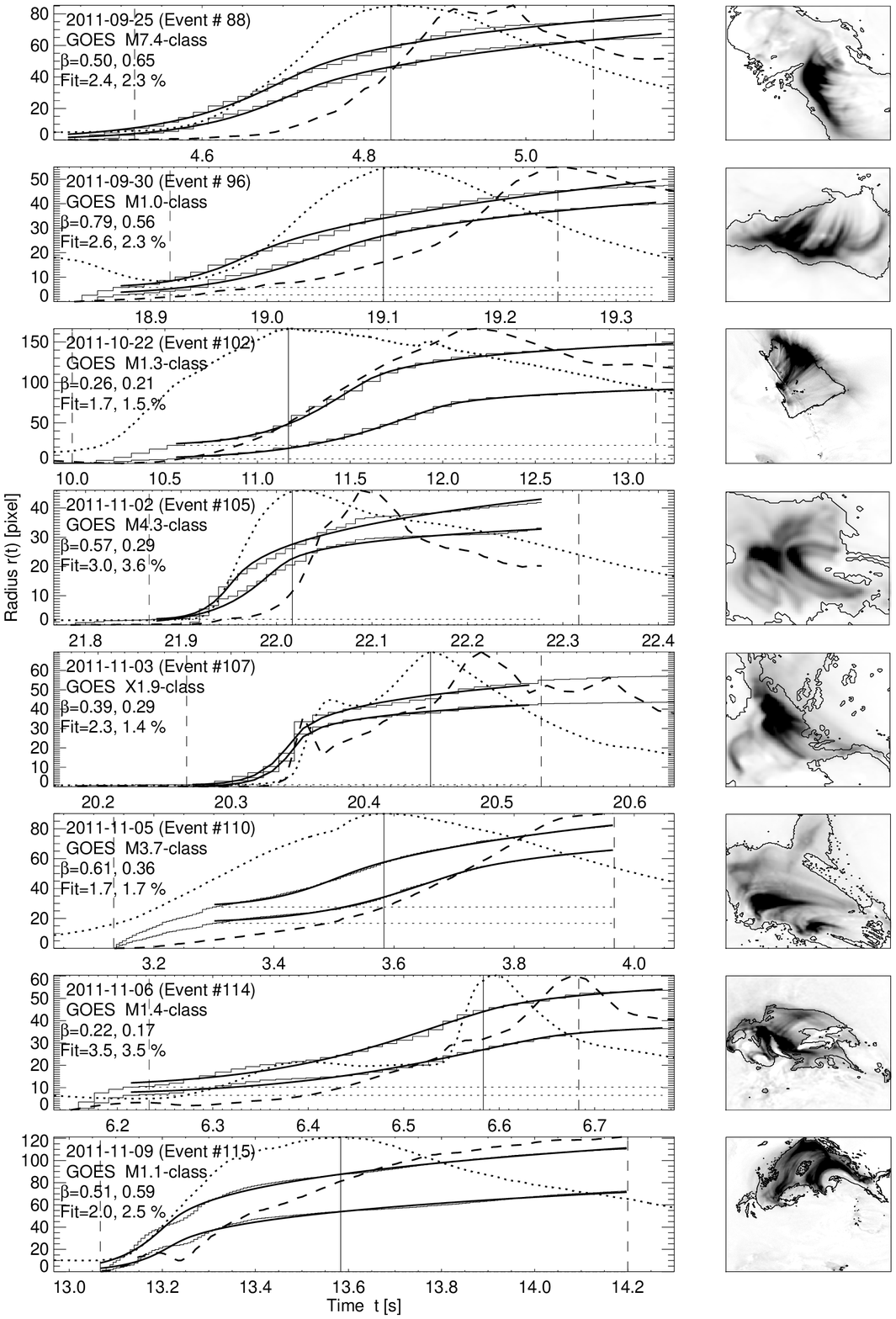}
\plotone{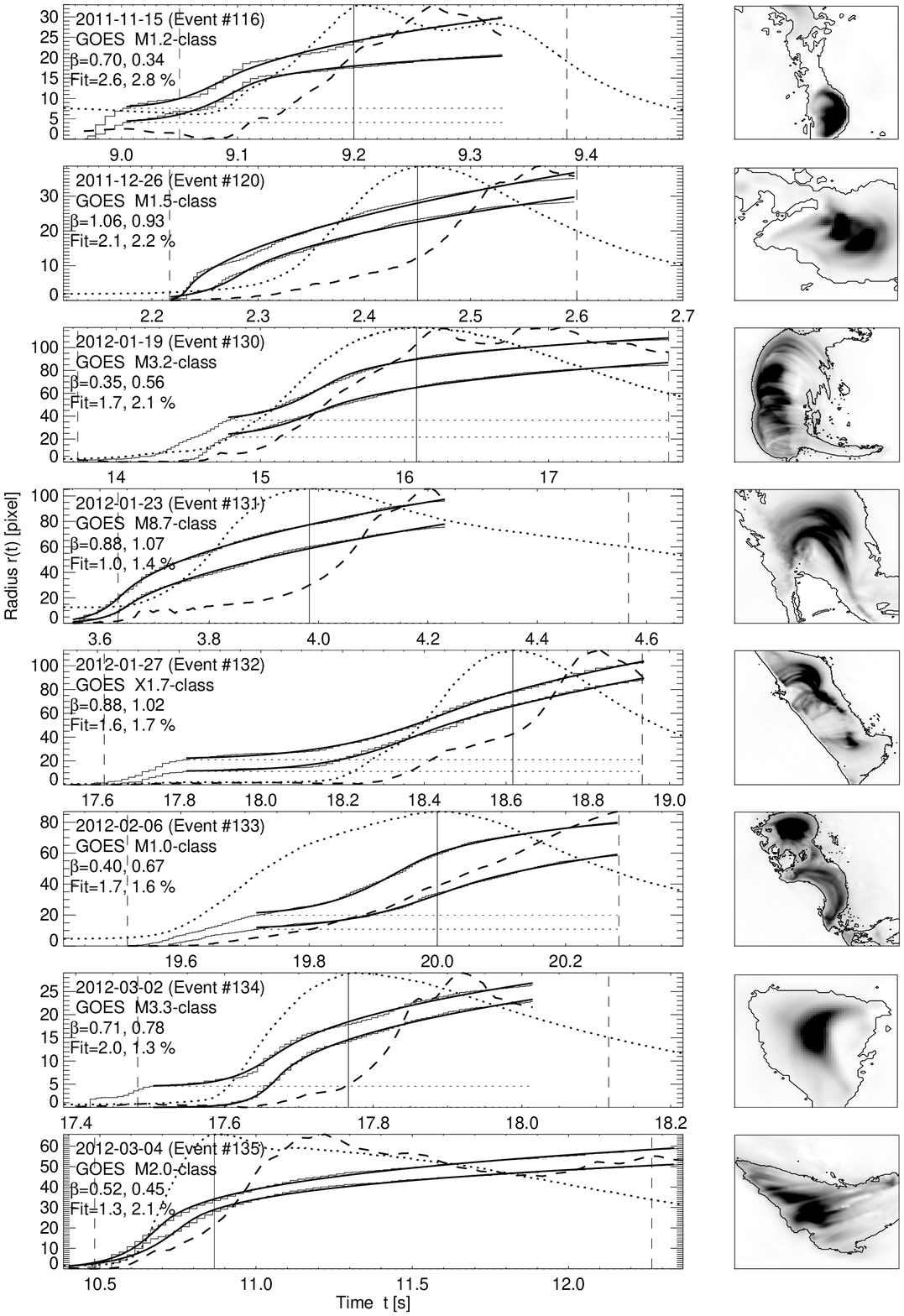}
\plotone{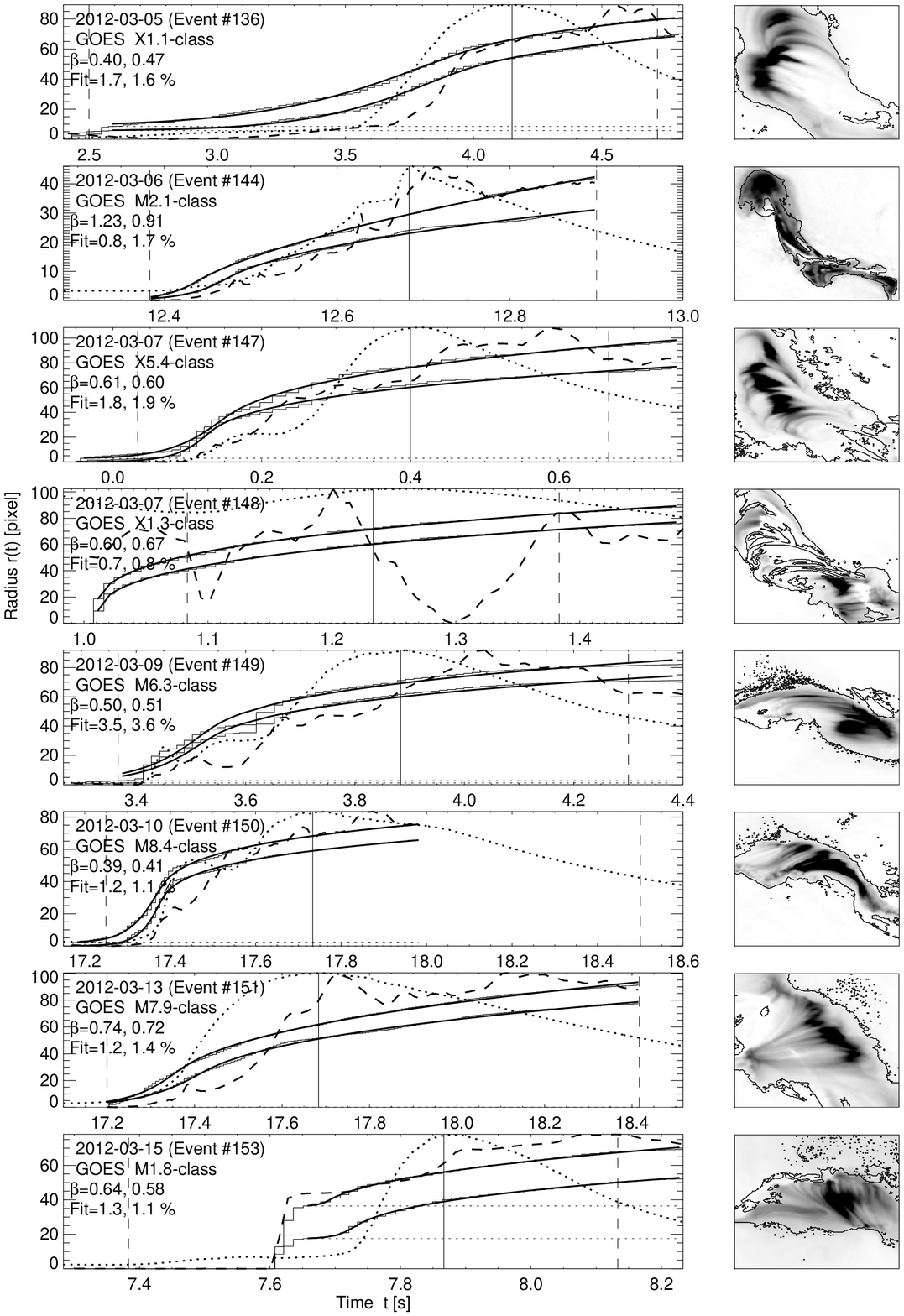}

\begin{figure}
\plotone{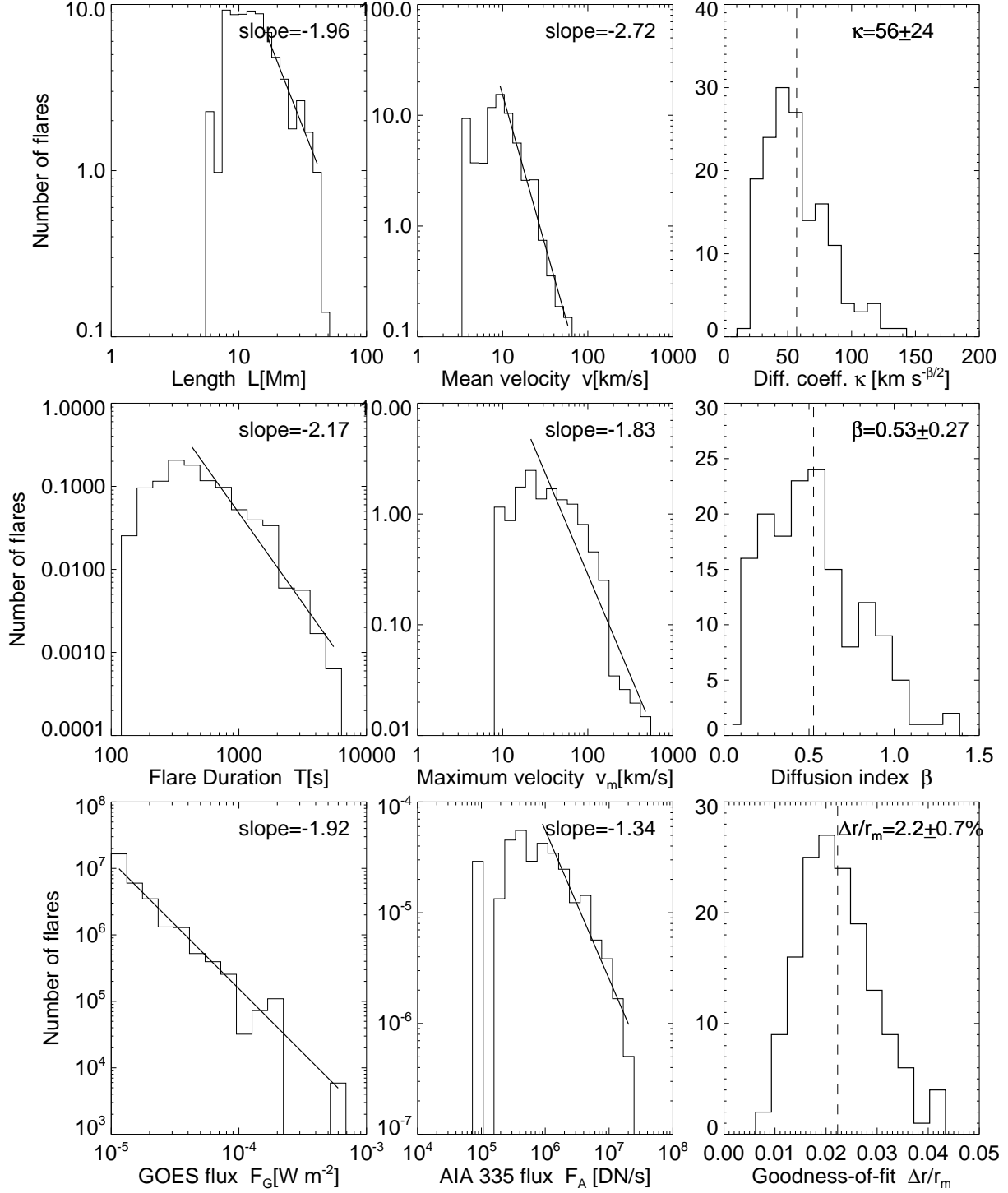}
\caption{Distributions of various observed parameters $(L, T, F_G,
v, v_m, F_A)$ in log-log format with powerlaw fit (left and middle column),
and best-fit model parameters $(\kappa, \beta, \Delta r/r_m$) with linear
histogram and mean and standard deviation indicated (right column).}
\end{figure}

\begin{figure}
\plotone{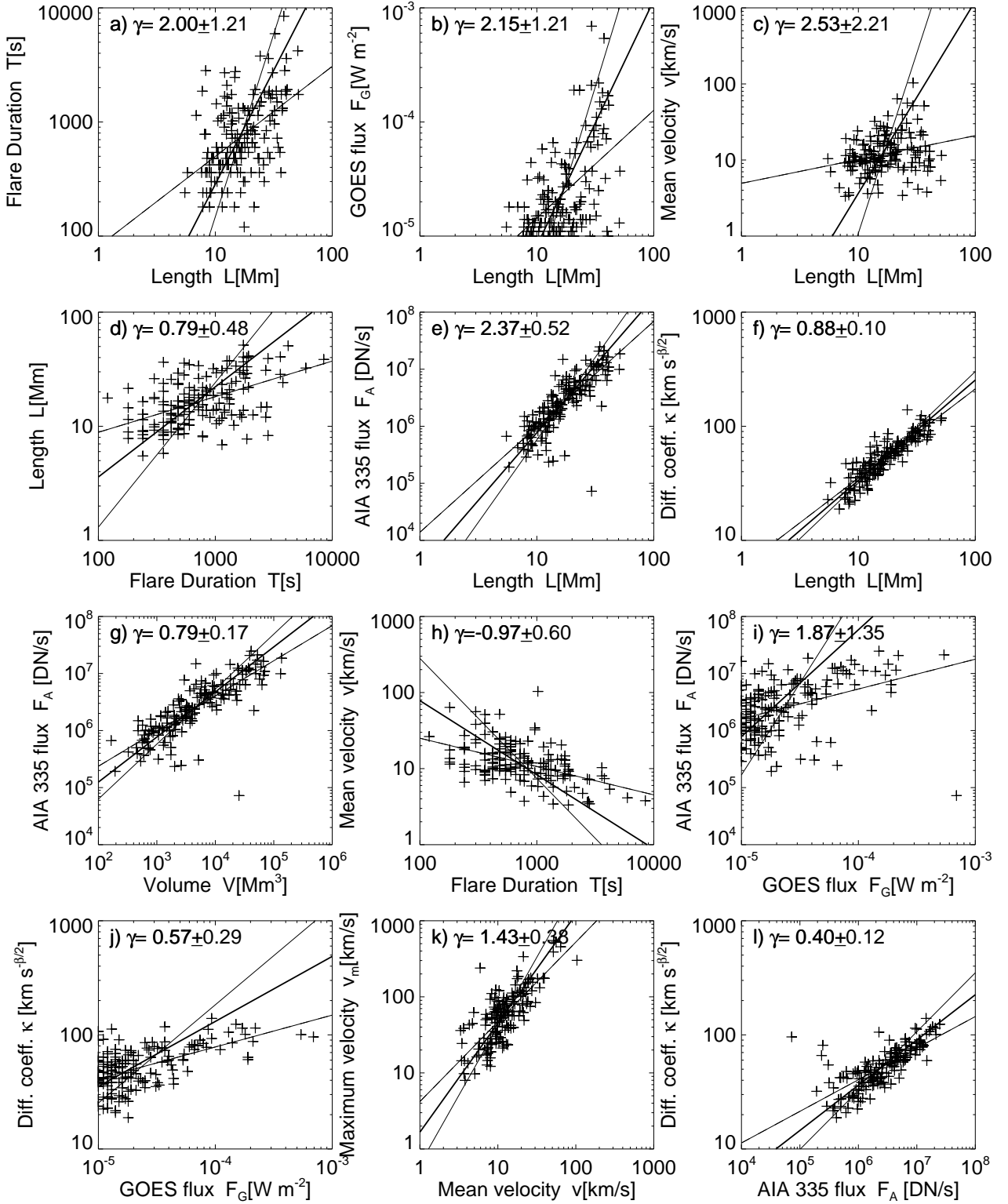}
\caption{Correlations between observed and best-fit parameters. The linear
regression fits are shown for $y(x)$ and $x(y)$ (thin lines), and for the
arithmetic mean (thick lines).}
\end{figure}

\end{document}